# Black holes at cosmic dawn in the redshifted 21cm signal of HI

I.F. Mirabel[1,2] and L.F. Rodríguez[3,4]

[1]Instituto de Astronomía y Física del Espacio (IAFE). Universidad de Buenos Aires. CC 67–Suc. 28 (C1428ZAA) CABA, Argentina

[2]Lab AIM, CEA/CNRS/Université Paris-Saclay, Université de Paris, F-91191 Gif-sur-Yvette, France

[3] Instituto de Radioastronomía y Astrofísica, Universidad Nacional Autónoma de México, P.O. Box 3-72, 58090, Morelia, Michoacán, México

[4] Mesoamerican Center of Theoretical Physics, Universidad Autónoma de Chiapas, 29050 Tuxtla Gutiérrez, Chiapas, México

**Abstract.** The first stars (Pop III stars) and Black Holes (BHs) formed in galaxies at Cosmic Dawn (CD) have not been observed and remain poorly constrained. Theoretical models predict that indirect insights of those Pop III stars and BHs could be imprinted as an absorption signal in the 21cm line of the atomic hydrogen (HI) in the cold Intergalactic Medium (IGM), against the Cosmic Microwave Background (CMB), when the Universe was less than 200 million years old. The first tentative observation of an HI absorption in the 21cm line at redshifts z > 15 by the Experiment to Detect the Global Epoch of Reionization Signature (EDGES) stimulated a great deal of research. To explain the additional large amplitude of the EDGES absorption signal a plethora of models have been proposed based on exotic physics and on astrophysical sources. Among the latter are those based on the hypothesis of an additional synchrotron Cosmic Radio Background (CRB) from BH-jet sources that boost the HI absorption signal at CD. The recent discovery of radio loud supermassive black holes (SMBHs) of $\sim 10^9 M_\odot$ in high-z quasars of up to z ~7 suggests the existence of a CRB component from growing BHs at z > 15, of unknown intensity.

To match the onset of the EDGES signal a CRB of comparable intensity to that of the CMB would be required. With no judgment on whether the EDGES signal is of cosmic origin or not, here we provide approximate calculations to analyze this type of absorption signals, taking that of EDGES as an example to explore what could be learned on BHs at CD from the onset of that type of HI signals. Assuming a BH mass to radio luminosity ratio as observed in radio-loud SMBHs of $\sim 10^9 M_\odot$ in quasars at z = 6 – 7, we find that rapidly growing radio luminous BHs of Intermediate Mass (IMBHs) of $10^{3-5} M_\odot$, in their way to become SMBHs, are the only type of astrophysical radio sources of a CRB that could explain the amplitude of the HI absorption reported by EDGES in the interval of z = 18 – 20. At those redshifts the EDGES signal would imply that the global mass density of IMBHs must be dominant over that of stars, more than 70% of the maximum Stellar Mass Density (SMD) expected at those high redshifts. This suggests that those IMBHs are formed before and grow faster than the bulk of stars, with no need of a large mass contribution from stellar-mass BH remnants of typical Pop III stars. The highly redshifted signals from these IMBHs at cosmic dawn may be detected at long radio wavelengths with the next generation of ultrasensitive interferometers such as the Square Kilometer Array (SKA), in the infrared with the James Webb Space Telescope (JWST), and in the X-rays with future space missions.

# I. Introduction

The so-called "Dark Ages" of the Universe start ≈380.000 years after the Big Bang, as matter cools down and space becomes filled with neutral hydrogen. The following Epoch of Reionization (EoR; Mesinger 2019) is a complex process that lasts up to about a billion years. It is believed that between the Dark Ages and the EoR there should be a phase existing over a relative short interval of time referred in the literature as the Cosmic Dawn (CD), produced by the appearance of Pop III stars, with the most massive ones promptly collapsing to BHs.

The proposed 21cm absorption of HI against the CMB is due to the Wouthuysen (1952) - Field (1958) effect (WF effect), a mechanism that couples the spin temperature of HI to the Ly$\alpha$ radiation from Pop III stars. The HI absorbs L$_{y\alpha}$ photons from Pop III stars, and then re-emits Ly$\alpha$ photons from either of the two spin states. This process causes a redistribution of the electrons between the hyperfine states of the HI ground state, which decouples the spin temperature of HI from the CMB, bringing it to the lower temperature of the IGM, allowing the HI to be observed in absorption against the CMB. Therefore, it is expected that the CD should be a phase over a relative short interval of time during which it may be possible to observe the HI in absorption against the CMB, before the gas is heated above the CMB temperature by astrophysical sources (Madau, Meiksin & Rees 1997 for an early review; Zaldarriaga, Furlanetto & Hernquist 2004; Furlanetto 2006; Furlanetto, Ho & Briggs 2006 for a review; Pritchard & Loeb 2010).

The detection of this absorption signal in the highly redshifted 21cm line of HI is one of the main scientific motivations of several ongoing radio astronomy projects at low frequencies. Interferometers such as SKA (Prandoni & Seymour 2015), HERA (De Boer et al. 2017), LOFAR (Shimwell et al. 2019), NenuFAR (Zarka et al. 2012), GMRT (Paciga et al. 2013), LOFAR (Yatawatta et al. 2013), PAPER (Parsons et al. 2010), LWA (Ellingson et al. 2009), and MWA (Bowman et al. 2013; Tingay et al. 2013) can map the three-dimensional fluctuations by tomography of the redshifted 21cm brightness temperature of the HI gas.

An alternative lower-cost and faster approach is the measurement of the global 21cm signal integrated over the sky by single dipole experiments such as SCI-HI (Voytek et al. (2014), EDGES (Bowman et al. 2018a), SARAS (Singh et al. 2018), LEDA (Price et al. 2018), PRIZM (Philip et al. 2019), REACH (Bevins et al. 2021) and the radiometer SARAS 3 (Singh et al. 2022).

In section II it is described the first tentative detection of a redshifted 21cm absorption of HI by EDGES and it is compared to an example of previous models. In section III we review the models proposed to explain EDGES, with particular emphasis in the one based on an additional CRB produced by relativistic jets from astrophysical BHs that boost the amplitude of the HI absorption. In section IV are discussed several possible caveats published in the literature to the hypothesis of a CRB of astrophysical origin. In section V it is estimated the intensity of the CRB and a general equation is derived to calculate the numbers of currently known types of radio sources that would be needed to match such CRB. In section VI we review BHs in binary stellar systems that radiate X-rays (BH-XRBs) as potential sources of the CRB, concluding that they cannot account for the intensity of the CRB required by EDGES, because they are inefficient radio sources, for physical reasons that are explained in the same section. In section VII we show that the only type of radio sources that can match the onset of EDGES are the radio luminous IMBHs progenitors at z =18-20 of SMBHs at z =

6-7. Assuming a BH mass to radio luminosity ratio as observed in the most distant radio loud quasar currently known, in the same section it is inferred that about more than 70% of the total expected maximum stellar mass density must be invested in IMBHs at those very high redshifts. In section VII it is also proposed that those IMBHs can be formed preferentially at CD, as transient BHs on their way to become SMBHs during the EoR, because of the very large column densities of gas that prevailed at those early epochs, and propose that this may explain the dearth of IMBHS in the local Universe. In section VIII we shortly review the observations of those large column densities, and in section IX we discuss the possibility to detect those IMBHs at CD with SKA. Finally, in section X the conclusions are presented by order of importance.

## II. A first tentative detection by EDGES

The report of a detection of HI absorption at z > 15 by EDGES (Bowman et al. 2018a) motivated interesting research on the potential implications of that type of HI signals from the early Universe. However, the EDGES signal caused surprise and some skepticism because it has an amplitude much larger than the amplitude predicted against the CMB alone, and has an extended bottom-flat shape rather than the more Gaussian shape predicted by models. The reported signal is about 3 orders of magnitude below the sky foreground and Hills et al. (2018) among others, expressed concerns about the analysis of the EDGES data. Those concerns were discussed in a reply by Bowman et al. (2018b), and it is clear that this controversy requires assessment by independent observations. Recently, after submission of this review, it was reported a non-detection of the EDGES signal by observations with the radiometer SARS 3, using a polynomial of $6^{th}$ order to fit the baseline of the data (Singh et al. 2022). Making no judgment on whether the EDGES signal is of cosmic origin or not, here we provide approximate calculations to analyze this type of absorption signals, taking that of EDGES as an example to explore what could be learned on radio loud BHs at CD from the onset of that type of HI absorption signals.

In Figure 1 is shown a comparison of the absorption reported by EDGES (in black) with an example of previous models of the predicted 21cm absorption against the CMB, that includes X-ray sources of astrophysical origin (in colors), such as accreting BHs, Supernovae (SNe) and massive X-ray binaries (e.g. Glover & Brandt 2003; Madau et al. 2004; Ricotti and Ostriker 2004; Furlanetto 2006; Pritchard & Loeb 2010; Glover 2013). Those X-ray sources would heat the gas (Glover & Brandt 2003; Mirabel et al. 2011; Mesinger, Ferrara and Spiegel 2013), decreasing the depth of the absorption, leading to a smoother end of the EoR (Haiman 2011).

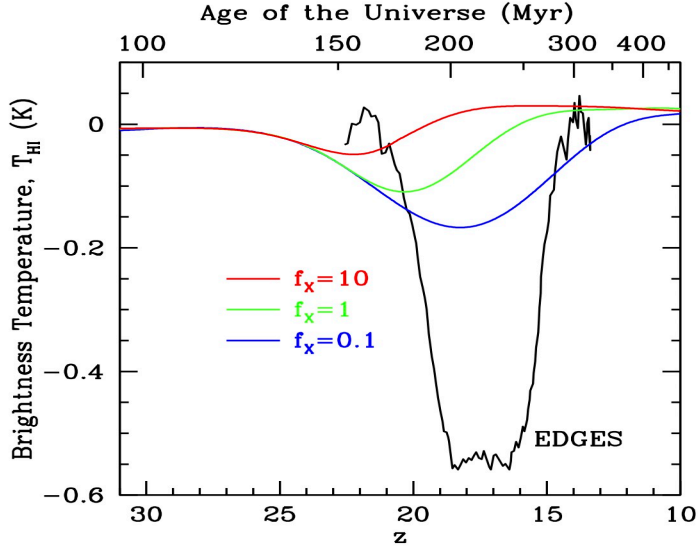

Figure 1: The black solid line shows the best-fitting 21-cm absorption profile for the global brightness temperature of an absorption centered at 78 MHz in the 60–99 MHz frequency band reported by the EDGES collaboration, plotted against redshift z and the corresponding age of the Universe (Bowman et al. 2018a). The color solid lines show a model example of the previously predicted brightness temperature absorption in the 21cm line of HI in the IGM against the CMB, averaged over the entire sky (Pritchard & Loeb 2010, here reproduced from Mirabel et al. 2011). The colors correspond to different possible values of $f_x$, which is a correction factor of the correlation between star formation and X-ray luminosity of galaxies in the local Universe (Gilfanov, Grim and Sunyaev 2004), that measures the heating of the HI by X-ray sources in galaxies at very high redshift (Furlanetto 2006). In the model example of this figure are shown the decreasing amplitudes and frequencies of the HI absorption with increasing X-ray heating of the IGM. For the lowest X-ray heating ($f_x \sim 0.1$ in this model example), the predicted frequency of the largest HI absorption amplitude against the CMB is at ~73 MHz, close to the 78 MHz reported by EDGES. Even in the most favorable case of this model for a large amplitude absorption against the CMB (blue curve), the predicted amplitude of the absorption signal is 2-3 times smaller than the amplitude of the EDGES signal.

## III. Possible explanations of the onset of the EDGES absorption: A Cosmic Radio Background (CRB) from BHs

To explain the large additional amplitude of the EDGES absorption, there have been proposed several new physics scenarios. One example is dark matter interaction with the gas that may cool it below the temperature of the IGM (Barkana 2018; Fialkov et al. 2018). In addition, there have been proposed several new physics scenarios that add an extra radio background that boosts the HI absorption against the standard CMB background. Such additional CRBs could be due to soft photon emission from light dark matter (Fraser et al. 2018); Rayleigh-Jeans tail emission of the CMB (Pospelov et al. 2018), radiative neutrinos (Chianese et al. 2018), and cosmic strings (Brandenberger et al. 2019).

On the other hand, there have been proposed explanations based on astrophysical agents. One is the possible non-previously accounted effects of the Lyα radiative background (Madau 2018; Kaurov et al. 2018; Meiksin & Madau 2021). Other astrophysical explanations are based on an additional CRB of astrophysical origin that boosts the HI absorption (Bowman et al. 2018a; Feng & Holder 2018; Chatterjee et al. 2020). In this context, there have been several models that do not constrain the origin of the CRB to specific astrophysical sources. For instance, Mirocha & Furlanetto (2019) parametrize the efficiency of the CRB by a factor $f_R=1$ as a function of the star formation rate (SFR) and the radio emission as observed in star-forming galaxies, finding that to match EDGES $f_R$ must be as extreme as 50. More speculative models of the CRB have been inspired by the ARCADE2 and LWA1 observations (Fialkov & Barkana 2019), or developed as a "state of the art" model based on a set of eight free parameters (Reis, Fialkov & Barkana 2020).

Two models identify and discuss the specific nature of the astrophysical sources that may produce the bulk of the CRB at CD: 1) supernovae (SNe) explosions of very massive stars at $z > 20$ (Jana, Nath & Biermann 2019), and 2) radio emission from a very high seeding fraction of efficiently accreting BHs (Ewall-Wise et al. 2018, 2020). Here we present a complementary study where we analyze in detail the properties of the following potential astrophysical sources of the CRB: SNe, SNe remnants (SNRs), stellar BHs in X-ray binaries (BH-XRBs), and SMBHs in high z quasars. By explicit back of the envelope calculations, and based on the observational knowledge of these sources, we show that luminous, rapidly growing massive BHs at CD are the only type of astrophysical radio sources that produce the bulk of the CRB that can match the onset of the HI absorption reported by EDGES.

The hypothesis of a synchrotron CRB that would explain the EDGES signal was first proposed in the context of the previous report of an all sky synchrotron radiation at frequencies below 1 GHz with the NASA balloon-borne instrument ARCADE 2 (Fixsen et al. 2011). This all sky synchrotron radiation was later confirmed between 40 and 80 MHz by all-sky maps with the LWA1 Low Frequency Sky Survey (LLFSS; Dowell and Taylor 2018). Observations at radio wavelengths (Condon et al. 2012) and in the far-infrared (Ysard & Lagache 2012) have respectively shown that resolved AGN-driven sources and star forming galaxies cannot provide the bulk of the radio synchrotron radiation observed with ARCADE 2, which could suggest that the residual from foreground sources may be due to a diffuse extragalactic background of emission that comes from very early cosmic times (Seiffert et al. 2011). To explain the additional amplitude of absorption reported by EDGES only a very small fraction of the ARCADE 2 background would be needed (Feng and Holder 2018). However, it should be noted that the interpretation of the ARCADE 2 measurement as an extragalactic contribution is still debated (Subrahmanyan and Cowsik 2013; and conference summary by Singal et al. 2018). In what follows we discuss possible caveats to the BH-CRB hypothesis.

## IV. Caveats to a CRB that can match the onset of EDGES

IV.1 Can the synchrotron radio emission from BHs be suppressed by the CMB at CD?

It has been argued that astrophysical radio backgrounds cannot explain the EDGES 21-cm signal due to constraints from Inverse Compton (IC) cooling of non-thermal electrons by the CMB, which would make astrophysical CRBs inefficient, in particular, the astrophysical scenarios based on the acceleration of relativistic electrons by accretion onto BHs and by supernovae (SNe) of Pop III stars (Sharma 2018). In what follows we show that this argument may be valid for the extended and diffuse radio emission in shock regions at large distances from the BH engine and Supernova Remnants (SNRs), where magnetic fields are at the most a few tens of μG. However, this argument is not valid for the BH compact synchrotron radio jets observed in accreting BHs of all mass scales, where magnetic fields are much larger.

X-ray Binaries (XRBs) that host stellar BHs are also called Microquasars (MQs) because they mimic in smaller scales the phenomena observed in quasars, namely a rotating black hole, an X-ray luminous accretion disk and synchrotron-emitting radio jets (Mirabel et al. 1992; Mirabel & Rodríguez, 1994, 1999). In the low/hard X-ray state these sources produce synchrotron radio emission of flat/inverted spectral index by quasi-steady, highly collimated compact jets with typical lengths of 10$\lambda_{cm}$ AU (Dhawan, Mirabel & Rodríguez, 2000). When referring to these compact jets in BH High Mass X-ray Binaries (HMXBs) we call these systems BH-HMXB-MQs.

As an example, we now calculate the minimum magnetic field B in the compact jet of Cygnus X-1, and the ratio of the IC cooling to the synchrotron cooling of that compact jet at CD. From the VLBI images (Stirling et al. 2001 & Miller-Jones et al. 2021) we estimate the major $\theta_{maj}$ and minor $\theta_{min}$ axes of the compact jet to have angular dimensions of 10 and < 2 milliarcsec (mas), respectively. Following the derivation of Condon & Ransom (2016), the minimum magnetic field of the jet is given by

$$B_{min} = [4.5(1+\eta)c_{12}]^{2/7}[\frac{L_r}{ergs^{-1}}]^{2/7}[\frac{\theta_{maj} \times \theta_{min}}{4mas^2}]^{-3/7}[\frac{1.5 \times 10^{13} d}{kpc}]^{-6/7},$$

where $\eta$ is the ion/electron energy ratio and $c_{12}$ is a constant that for a flat spectrum and a radio luminosity integrated up to 220 GHz (Fender et al. 2000) is given by $4 \times 10^6$. Assuming $\eta = 0$, it is obtained $B \geq 0.13G$.

For the ratio of the cooling time $t_{IC}$ of IC off the CMB to the synchrotron cooling time $t_{syn}$

$t_{IC}/t_{syn} \sim 0.91$ (B/$10^{-3}$G)$^2$ [(1+z)/18]$^{-4}$     (eq. 6 in Sharma 2018)

assuming a redshift z = 19 and B > 0.13G it is found $t_{IC}/t_{syn} > 10^4$. This implies that the synchrotron cooling of the compact jets dominates (they release radio emission), well before they could be suppressed by IC cooling off the CMB at cosmic dawn.

The synchrotron lifetime is given by (Condon & Ransom 2016):

$\tau_S \simeq c_{12} B^{-\frac{3}{2}} s,$   that for $B \simeq 0.13G$ gives

$\tau_S \simeq 8.5 \times 10^7 s = 2.8 yr.$

Although relatively brief, this timescale is much larger than the time required for relativistic ejecta to move across the detectable semi-major axis of the jet. This semi-major axis is about 5 mas or $1.7 \times 10^{14} cm$, that can be traveled by ejecta moving close to the speed of light in only $5.6 \times 10^3 s$.

We conclude that the synchrotron cooling of the electrons in MQ compact radio jets dominates over IC cooling produced by the CMB photons by more than four orders of magnitude, and the radiation from these compact BH-HMXB-MQ jets at 21cm is not suppressed by the CMB photon densities, even at redshifts of z~20. Since the typical main sequence lifetimes of massive donor stars in BH-HMXBs are of ~$10^5$yr, the synchrotron losses in the compact jets of MQs may be compensated by the quasi-steady supply of fresh relativistic electrons during BH accretion times of > $10^5$yr, the typical lifetime of the massive donor stars. On the contrary, the magnetic fields in the large-scale bow shocks produced by jets from compact sources and SNRs are at most of tens of μG, which make them inefficient radio sources to account for the bulk of the CRB needed to explain the onset of the HI absorption reported by EDGES.

On the much larger scales of SMBHs the importance of magnetic fields in twin compact jets of AGNs and QSOs has been highlighted by Zamaninasab et al. (2014), and it has already been noted that the radio emission that arises in the inner, strongly magnetized compact core of SMBH jets should not be significantly affected by IC scattering off CMB photons (Ghisellini et al. 2014). If the minimum magnetic fields in radio emitting cores of AGN are in the range of B = $10^2$–$10^4$ G (e.g. Baczko et al. 2016), the synchrotron cooling of compact jets in AGN at 21cm dominates over IC cooling by the CMB photons at z~20 by orders of magnitude of up to ~$10^9$.

IV.2 Are X-rays from BHs an obstacle to match the onset of the EDGES absorption?

The radio emission from accreting BHs is observed to be coupled with X-rays in real time in stellar BHs (e.g. Mirabel et al.1998) and Supermassive BHs (SMBHs; e.g. Marscher et al. 2002). Universal radio–X-ray correlations have been found for BH-XRBs in the low/hard X-ray state (Gallo et al. 2005; 2018), and in BHs of all mass scales (Merloni et al. 2003; Falcke et al. 2004). In AGN the radio 20 cm and hard X-rays (20-100 keV) are strongly correlated, which supports the idea that in efficiently accreting BHs these two physical components (X-rays and radio) are produced as part of the same process, or are causally connected (Malizia et al. 2020).

In this context, the radio emission from accreting BHs needed to match the EDGES absorption signal would be coupled with X-ray emission that heats the gas, reducing the amplitude of the HI absorption signal with increasing $f_x$ (as it is shown in the model example of Figure 1), unless the potentially heating soft X-rays of < 1 keV produced by the accreting Pop III BHs are locally blocked by high density H+He pristine gas (Wilms, Allen & McCray 2000), and do not reach the cold HI of the IGM to large distance scales. Therefore, current models of the EDGES absorption based on a BH-CRB propose that the radiation at radio wavelengths from Pop III BHs is far more efficient than heating by X-rays (Mebane et al. 2020), and this may be the case if Pop III BHs are embedded in high column density gas of $N_H$ ~ 5 x $10^{23}$ at $cm^{-2}$ (Ewall-Wice et al. 2018; 2020).

We note that such local large column densities of gas are observed in Radio Luminous BH-XRBs as SS 433 (Fabrika 2004; Fabrika et al. 2017; Cherepashchuk et al. 2020), V404 Cyg (Miller-Jones et al. 2019) and GRS 1915+105 (Motta et al. 2021), during super-Eddington accretion of large X-ray absorption/obscuration with associated activity of radio jets. The typical intrinsic X-ray covering absorption in these stellar BH jet sources are by local column density gas of $N_H$ ~ $10^{23-24}$ atoms $cm^{-2}$. In the HMXB Cygnus X-3 the low-energy X-ray cut-offs are due to a wind column density of ~$10^{23}$ at $cm^{-2}$ from a donor Wolf-Rayet star (Koljonen & Tomsick 2020) in orbital period of 4.8h with the compact object.

In SMBHs are also found such large local column densities of gas. New population synthesis models suggest that at least 50% of AGNs up to z=0.1 are Compton thick, which implies gas column densities $N_H$ > $10^{24}$ $cm^{-2}$ (e.g. Anhanna et al. 2019). Such local and even larger column densities of gas are observed in high redshift quasars, in particular where the SMBHs may be efficient radio sources with the soft X-rays locally absorbed.

IV. 3 Are the CMB, CRB and Lyα emission obstacles to produce the required absorption?

In the absence of X-ray heating, Venumadhav et al. (2018) have pointed out that heating from the CMB photons will produce at z = 17 a 10% increase in the kinetic temperature of the gas.

In the case of an additional CRB this heating is expected to increase accordingly. However, it should be noted that for the z range we analyze here, namely, the onset of the EDGES absorption that takes place between z= 20.5 and z= 18.5, the CMB effect is much smaller, between 1% and at most 5% (see Figure 3 of Venumadhav et al. 2018). Then, we do not take this effect into account in our approximate calculations.

Another subtle heating effect first noted by Chuzhoy & Shapiro (2007) may be due to Lyα emission from stars. This heating impact of Lyα emission on the global 21-cm signal has recently been analyzed together with that of the CMB by Reis, Fialkov and Barkana (2021). For $f_X = 0$ (no X-rays) these authors show that the difference in gas temperature at the onset of the EDGES absorption (z = 18.5 to 20.5), with and without Lyα and CMB is less than ~10% (see top right of Figure 4 by Reis et al. 2021). When including these effects the models by these authors reach an HI absorption floor at z ~19 with a maximum absorption depth of −165mK (see bottom left of Figure 4 by Reis et al. 2021). This maximum absorption depth of −165mK with no CRB is similar to that in the model example by Mirabel et al. (2011) reproduced in Figure 1.

Here we do not aim to present a new detailed model of the EDGES absorption in the whole redshift interval of z=13-21. We only aim to discuss the possible nature of the BH-jet sources of a CRB that could match an onset of the absorption in the interval of z = 18.5 to 20.5 as that of EDGES. Therefore, in our approximate calculations we ignore subtle, second order effects due to Lyα radiation, CMB and CRB on the gas temperatures at CD that together amount to less than 10%.

## V. The CRB that matches the onset of the EDGES absorption

### V.1 The required intensity of the CRB

In our approximate calculation of the required intensity of a BH-CRB to match the onset of the EDGES trough in the z =18.5-20.5 range we assume $f_X = 0$ (no X-rays) and follow the theoretical framework by Pritchard and Loeb (2008).

In the absence of heating, the brightness temperature at 1.42 GHz and adjacent frequencies taking into account only the spin temperature $T_S$ and the cosmic background temperature $T_{CMB}$ is given by

$$T_{HI}(\nu) = (T_S - T_{CMB})(1 - e^{-\tau_\nu}),$$

where $\tau_\nu$ is the optical depth of $HI$ at frequency $\nu$. If there is an additional cosmic radio background with brightness temperature $T_{CRB}$ the absorption becomes deeper as indicated by the modified transfer equation:

$$T'_{HI}(\nu) = (T_S - T_{CMB} - T_{CRB})(1 - e^{-\tau_\nu}).$$

We then have that the ratio of the absorption with an additional cosmic radio background (indicated with a prime symbol) to the absorption without it is given by:

$$\frac{T'_{HI}(\nu)}{T_{HI}(\nu)} = 1 + \frac{T_{CRB}}{T_{CMB} - T_S}$$

from where the brightness temperature of the additional cosmic radio background is given by:

$$T_{CRB} = [\frac{T'_{HI}(\nu)}{T_{HI}(\nu)} - 1](T_{CMB} - T_S).$$

Comparing in Figure 1 the peak amplitude of the absorption in the EDGES profile with that in the model example of the HI absorption against the CMB where soft X-rays (< 1keV) from BHs would have a relative small heating effect on the gas (the most favorable case for $f_X = 0.1$ in this model indicated by the blue curve in Fig. 1), we have that

$$\frac{T'_{HI}(\nu)}{T_{HI}(\nu)} \simeq 3.4.$$

The cosmic background temperature is given by $T_{CMB} = 2.7(1 + z)$, so that at $z = 19.5$ we have that $T_{CMB} \simeq 55K$. Finally, at $z \simeq 19.5$, we have that $T_S \simeq 20K$ (Pritchard & Loeb 2008).

Then, to explain the deep absorption reported by EDGES we require an additional radio background $T_{CRB} \simeq 85K$, about 1.5 times the $T_{CMB}$ at that epoch. At present $T_{CMB} = 2.7K$ and we then need $T_{CRB} \simeq 3.9K$ at the frequencies of the EDGES detection.

In the following the calculations are given in the units used in radio astronomy and therefore is useful to express this $T_{CRB} \simeq 3.9K$ in Janskys (Jy):

The flux density of the cosmic microwave background per unit of frequency range is given by:

$$S_\nu = B_\nu \Omega,$$

where $B_\nu$ is Planck's function and $\Omega$ is the solid angle considered. Integrating over all the celestial sphere, $\Omega = 4\pi$, and

$$S_\nu = 4\pi B_\nu.$$

Then, for a temperature of $T_{CMB} = 2.7K$ and a frequency of $\nu = 73$ MHz, that corresponds to

$$S_\nu(T_{CMB}) = 5.54 \times 10^{-20} erg s^{-1} cm^{-2} Hz^{-1} = 5.54 \times 10^3 Jy$$

and for the $T_{CRB}$ we obtain

$$S_\nu(T_{CRB}) = 8.31 \times 10^{-20} erg s^{-1} cm^{-2} Hz^{-1} = 8.31 \times 10^3 Jy \quad \text{eq. (1)}$$

V.2 The equation for the number of radio sources needed to match EDGES at CD

For this calculation we assume the ideal case where no soft X-rays reach the cold IGM ($f_X = 0$; Ewall-Wise et al. 2018; 2020). Assuming a quasi-steady mean radio flux density at 1.4 GHz $S_{1.4}$ of flat spectrum across the centimeter range, for a distance d the luminosity per frequency interval of the source is

$$L_{1.4} = 4\pi d^2 S_{1.4}$$

which remains valid not only at 1.4 GHz but also at adjacent higher and lower frequencies as well. At cosmological distances, such a source will produce a flux density at present given by (Hogg 1999):

$$S_\nu = (1+z)\frac{L_{1.4}}{4\pi D_L^2},$$

where $D_L$ is the luminosity distance (Weedman 1986; Ryden 2016) and $\nu$ is the redshifted frequency where the observation is made. The differential of flux density produced at present by a thin spherical shell at a given $z$ and with a thickness $\Delta z$ will be given by:

$$dS_\nu = (1+z)\frac{nL_{1.4}}{4\pi D_L^2}dV,$$

where $n$ is the number density of sources at $z$ and $dV$ is the differential of volume within $dz$. Then, the total flux density coming from all the sky and produced by a population of sources between $z_{max}$ and $z_{min}$, the redshift limits where the sources are present, will be given by:

$$S_\nu = \int_{z_{min}}^{z_{max}}(1+z)\frac{nL_{1.4}}{4\pi D_L^2}dV.$$

The luminosity distances and volumes are taken from Wright (2006) for a flat Universe (Ned Wright's Javascript Cosmology Calculator).

If we assume that the formation of radio sources is limited to the range of $z$ =20.5 to $z$ =18.5 we can approximate the integral by evaluating the variables at the mid-point of this range, that is, at $z$ =19.5. We then have

$$S_\nu = (1+z)\frac{L_{1.4}}{4\pi D_L^2}\int_{18.5}^{20.5}ndV.$$

The total number of sources $N$ between $z$ =20.5 and $z$ =18.5, is given by

$$N = \int_{18.5}^{20.5}ndV$$

The equation is then given by

$$S_\nu = (1+z)\frac{L_{1.4}}{4\pi D_L^2}N.$$

In this equation the luminosity distance has to be in cgs units (centimeters) to be consistent with the characteristic luminosity per frequency interval of the individual sources. For $z$ =19.5, $D_L$ = 224.5 Gpc = 6.94×10$^{29}$ cm. We then in general have:

$$N = \frac{4\pi D_L^2 S_\nu}{(1+z)L_{1.4}} \qquad \text{eq. (2)}$$

that we use to calculate the numbers of different types of radio sources needed to match the onset of the EDGES absorption between z = 20.5 and 18.5.

## VI. Stellar Black holes in High Mass X-ray Binaries (BH-HMXBs)

VI.1 BH-HMXBs at cosmic dawn

The increase of computational power in recent years allowed more advanced simulations where due to fragmentation, Pop III stars of > 140 $M_\odot$ would not be as frequently formed as thought before, the majority of Pop III stars being formed with lower masses in binary and larger multiple systems (Bromm, Coppi & Larson 2002; Turk, Abel & O'Shea 2009; Stacy, Greif & Bromm 2010; Bromm & Yoshida 2011; Stacy, Bromm & Lee 2016; Sugimura, Matsumoto & Hosokawa 2020, among many others).

On the other hand, extensive observational campaigns on massive stars in the Milky Way and Large Magellanic Cloud (LMC) with the Hubble Space Telescope and new generation of ~10m ground-based telescopes have shown that the majority of massive stars are formed nearly exclusively in multiple systems, and if corrected for observational biases, the true multiplicity fraction might come close to 100%, despite different environments, sample ages and metallicities (Sana et al. 2012, 2014 & 2017; Mahy et al. 2020). Massive stars in the LMC that were believed to be of ~$10^3$ $M_\odot$ or more, with those more powerful telescopes have been dissected in multiple systems that contain only very few massive stars up to masses of ~220 $M_\odot$ (Bestenlehner et al. 2020; Schneider et al. 2021), that probably result from mergers of lower mass stars. There is evidence that such mergers may be taking place in the present Universe at sites of massive star formation (Zapata et al. 2017).

The Pop III stellar IMF is a complicated function of mass and despite the theoretical progress made, at present there is no definitive consensus of what that IMF at CD may be. Here it is assumed that at zero metallicity the stellar IMF is top heavy, and stars of 30 $M_\odot$ to 140 $M_\odot$ end as BHs by direct collapse or through failed SNe, rather than as energetic core-collapse SNe (Heger et al. 2002; Gregory et al. 2009, among others). Under this assumption primordial binaries with stars of > 30 $M_\odot$ end as BHs gravitationally bound with massive donor stars (BH-HMBs), which for sufficiently low orbital periods and eccentricities are persistent BH-HMXB-MQs.

From theoretical models (Heger et al. 2003) and observations in the local Universe (e.g. Fragos et al. 2013; Basu-Zych et al. 2013; Kaaret 2014; Jeon et al. 2014; Douna et al. 2015, 2018; Brorby et al. 2016) it is expected a metallicity dependence of the formation rate of BHs and HMXBs in galaxies, with higher formation rates of HMXBs in the early Universe than in the local Universe. In fact, from the Chandra South deep field an empirical relation of $L_{2-10\,keV}$ (HMXB)/SFR $\propto$ (1 + z) up to z=4 has been inferred (Lehmer et al. 2016). This observational result is consistent with predictions of population-synthesis models that attribute the increase of HMXB scaling relations with redshift as being due to the declining host galaxy stellar ages and metallicities (Lehmer et al. 2016). In these contexts, it would be expected a formation rate of BH-HMXBs at CD larger than in the local Universe.

In addition, theoretical models predict that massive Pop III stars up to 140 $M_\odot$ before ending directly as BHs, go through phases of Pulsation Pair Instability Supernovae (PPISN or "SN impostors"), that produce the ejection of the outer stellar layers (Heger & Woosley 2002; Heger et al. 2003; Woosley & Heger 2021). The mass lost by a PPISN alone can be of several tens of solar masses. Examples of PPISN may be the Eta Carina outbursts and the sudden disappearances from sight of very luminous blue variable (LBV) massive stars in nearby galaxies (Kochanek et al., 2008; Reynolds et al. 2015). One of those sight disappearances of luminous stars has recently been reported in PHL 293B, one of the most metal-poor known galaxies (Allan, Groh, Mehner et al. 2020).

These phenomena are interpreted as either due to the sudden obscuration by PPISNe mass ejection by surviving massive stars, or to the final formation of a stellar BH mass of up to 50 $M_\odot$. This is the lower boundary mass of a predicted BH mass gap between 50 $M_\odot$ (70 $M_\odot$ for Pop III stars in Woosley 2017) and 130 $M_\odot$ (Heger et al. 2003), where very few BHs should exist in close binaries that merge in a Hubble time. The discovery of a single merging BH pair in which the more massive component has a mass near 85 $M_\odot$ (Abbott et al. 2020) may be explained by several factors (Woosley & Heger 2021), which affect the theoretical estimate for the low boundary of the high BH mass gap. Recently, it has been confirmed that the mass distribution of a binary's more massive component strongly decreases as a function of primary mass, but there is no evidence of a strongly suppressed merger rate above ~60 $M_\odot$ from the GWTC-3 latest observing run (Ligo, Virgo & Karga Collaboration 2021).

How important could be the very early metallicity enhancement of the gas at CD by PPISN of Pop III stars, before the heavy elements injection by core collapse SN explosions, is an interesting, still open question.

VI.2 Cygnus X-1 as template of BH-HMXBs

In the local universe only three dynamically confirmed persistent BH-HMXBs sources of X-rays are known at present: Cygnus X-1, LMC X-1 and M33 X-7. All three have luminosities $L_X \sim 10^{38-39}$ erg sec$^{-1}$, binary orbital periods in the range of 3.5-5.6 days with eccentricities < 0.0256, BH masses in the range of 10 $M_\odot$ to 21 $M_\odot$, and massive donor stars of main sequence spectral type O9.7Iab in Cygnus X-1 and O-type more evolved supergiants in LMC X-1 and M33 X-7 (Orosz et al. 2011; 2009; 2007, respectively). In external galaxies, BH-HMXBs may be one type among the stellar compact sources observed as Ultra Luminous X-ray sources (ULXs; Feng & Soria 2011; Soria et al. 2021).

For the analysis of the role of BH-HMXB-MQs at CD we chose Cygnus X-1 as a local analog of this type of sources at that epoch. After the discovery of Cygnus X-1 as a luminous X-ray source (Bowyer et al. 1965), it was detected as a radio source (Braes & Miley 1971), and the precise position allowed its identification as the binary system that hosts the first BH discovered in the Universe (Webster & Murdin 1972; Bolton 1972). Since then, Cygnus X-1 has been the best studied BH-HMXB for about 50 years.

It has been proposed that the BH in Cygnus X-1 was formed in situ by direct/failed SN collapse of a > 40 $M_\odot$ star of the Cygnus OB3 association of massive stars (Mirabel & Rodrigues 2003). Recent VLBI radio astrometry of Cygnus X-1 (Miller-Jones et al. 2021), and optical astrometry of Cygnus OB3 stars with the Gaia satellite (Gaia collaboration 2018), confirmed that Cygnus X-1 is –within the errors– at the same distance of ~2.2 ± 0.18 kpc of

Cygnus OB3, concluding that Cygnus X-1 is a BH of 21.2 ± 2.2 $M_\odot$ accreting from a solar metallicity donor star of ~41 $M_\odot$ (Miller-Jones et al. 2021).

Since first detected at radio wavelengths, several radio monitoring programs with different radio telescopes by different teams have shown that the radio counterpart of Cygnus X-1 is a quasi-steady radio compact source of 10-20 mJy of flat/inverted spectrum for about 90% of the time. From 1550 daily measurements reported in the Green Bank Interferometer (GBI) survey of the National Radio Astronomy Observatory (https://en.wikipedia.org/wiki/Green_Bank_Interferometer), we infer that the radio source in the hard X-ray state has a mean quasi-steady flux at 1.4 GHz of 14.6 ± 0.5 mJy, which is consistent with measurements by several other teams. Fender et al. (2000) emphasized the remarkable flatness of the quiescent Cygnus X-1 spectrum up to 220 GHz. Assuming a constant flux density of 14.6 mJy between 1.4 and 220 GHz and a distance d = 2.2 kpc (Miller-Jones et al. 2021), it is inferred a radio luminosity of $L_r = 1.8 \times 10^{31}$ erg/s.

In Figure 2 (Left) reproduced from Gallo et al. (2005), it is shown a ring structure of ~5 pc radius that results from a strong shock at the location where the pressure exerted by the collimated milli-arcsec scale compact jet shown in the inset (Miller-Jones et al. 2021) takes place on the interstellar medium. To produce the shock ring structure it has been estimated a jet injection time of 0.02–0.32 Myr, a power $10^{36} < P_{jet} < 10^{37}$ erg s$^{-1}$, and a total injected energy of ~$10^{49}$ erg, with the bulk of the liberated accretion power in the form of 'dark', radiatively inefficient feedback at radio wavelengths (Gallo et al. 2005).

In Figure 2 (right) is shown the high energy spectrum of Cygnus X-1 observed with the INTEGRAL/IBIS space telescope. If the highly polarized gamma-ray emission of 76% ± 15% in the 400-keV to 2-MeV band reported from observations of Cygnus X-1 (Laurent et al. 2011; Jourdain et al. 2012; Rodriguez et al. 2015) has a compact jet origin, the compact jets likely originate at ~$10^3$ gravitational radii from the BH, with very efficient particle acceleration and a highly ordered magnetic field of $B > 10^4$ G (Zdziarski et al. 2012).

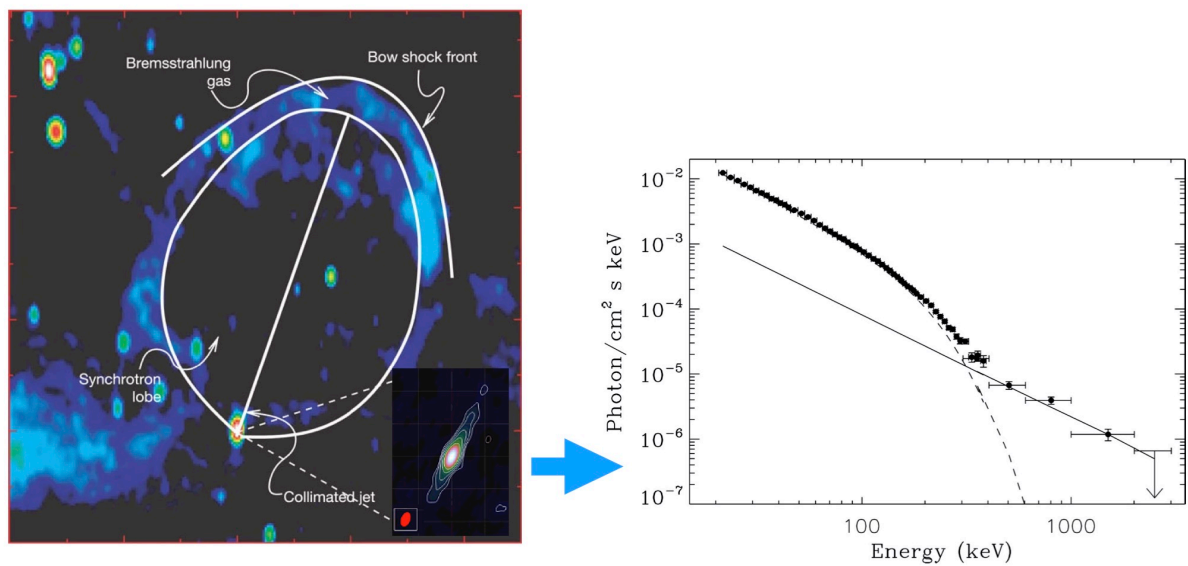

Figure 2. (Left) Sketch of the Cygnus X-1 model of a ~$10^6$ Astronomical Units (AU) diameter ring imaged at 21cm continuum with the Westerbork Synthesis Radio Telescope (Gallo et al. 2005). The ring is the result of a strong shock that develops at the location where the mechanical energy triggered by the collimated mas-scale

compact jet imaged by Miller-Jones et al. (2021; shown in the inset at a larger scale), strikes the ambient interstellar medium. The jet image was obtained with the Very Long Baseline Array of NRAO at 8.4 GHz. (Right) High-energy spectrum of Cygnus X-1 as measured by the INTEGRAL/IBIS telescope. Two components are clearly seen: a "Comptonization" spectrum caused by photons up-scattered by inverse Compton scattering off thermally distributed electrons in a hot plasma (dashed line), and a highly polarized component of 76% ± 15% at >400 keV (solid line), believed to be produced by the compact jet shown in the inset of the left-hand panel. The spectrum is reproduced from Laurent et al. (2011).

From the observations of the three dynamically confirmed BH-HMXBs it is concluded that this type of sources dissipates a large fraction of the liberated accretion power in the form of inefficient radio radiative relativistic outflows, rather than locally in the X-ray emitting region. A large fraction of the accretion power in these BH-HMXBs is converted into mechanical power that produce the shock ionized bubbles associated with some ULX sources in nearby star forming galaxies (Feng & Soria 2011; Soria et al. 2021). In what follows we discuss why BH-HMXBs may be luminous X-ray sources but dim radio sources.

VI.3 Why BH-HMXBs are inefficient at producing radio emission

Our analysis of Cygnus X-1, one prototypical present-day BH-HMXB, indicates it is a comparatively faint radio source. If we assume to first approximation that the radio emission is proportional to the mass of the BH and take as a reference source the supermassive BH P172+18 (Bañados et al. 2018) listed in Table 1 below, we would expect Cygnus X-1 to show a 1.4 GHz flux density of $1.2 \times 10^5$ Jy, about 8 orders of magnitude larger than observed.

This inefficiency of radio emission of BH-XRBs for a given BH mass when compared with supermassive BHs (SMBHs) has been noted by Merloni, Heinz & Matteo (2003), but to our knowledge has remained unexplained until present. We show here that this BH radio inefficiency in X-ray binaries is related to the presence of a companion star to the BH. For the sake of concreteness, we discuss this hypothesis taking Cygnus X-1 as the example.

A cooling mechanism of the jets that is generally not considered is that produced by inverse Compton scattering of the photons from the donor star on the relativistic electrons of the jet. In the case of sources like Cygnus X-1, where the donor star is very luminous, this mechanism can be very important. The energy density of the magnetic field in the jet is given by

$$u_B = \frac{B^2}{8\pi},$$

and since in Cygnus X-1 as determined before $B \simeq 0.13 G$ gives

$$u_B = 6.7 \times 10^{-4} erg\, cm^{-3}.$$

On the other hand, the photon energy density produced around a discrete source of radiation is given in cgs units by (Pittard et al. 2006):

$$u_R = \frac{L}{4\pi d^2 c}.$$

where $L$ is the luminosity of the source, $d$ is the distance between the source and the point considered and $c$ is the speed of light. In the case of Cygnus X-1, $d = 0.2 AU = 3.0 \times$

$10^{12} cm$ and $L = 3 \times 10^5 L_\odot = 1.1 \times 10^{39} erg s^{-1}$ for the donor stars of main sequence spectral index 09.7Iab as in Cygnus X-1. We then obtain

$u_R = 3.2 \times 10^2 erg cm^{-3}$.

As had been discussed before, the synchrotron lifetime is given by (Condon & Ransom 2016):

$\tau_S \simeq c_{12} B^{-3/2}$    that for $B \simeq 0.13 G$ gives

$\tau_S \simeq 8.5 \times 10^7 s = 2.8 yr$.

This means that the timescale for IC cooling is

$\tau_{IC} \simeq \dfrac{u_B}{u_R} \tau_S \simeq 180 s$.

This timescale is much shorter than the previously calculated time of $5.6 \times 10^3 s$ required for the ejecta to travel across the observed length of the jet at a speed close to that of light. Therefore, IC cooling by the photons of the donor star will affect the microscopic motions of the radiating relativistic electrons, suppressing the synchrotron emission. However, it will not affect the bulk motion of the ejecta. This is why BH-HMXB-MQs are capable to produce energetic mechanical feedbacks with radio "dark" jets.

VI.4 Similarities and differences between Cygnus X-1 and Pop III BH-HMXBs

Pop III BH-HMXBs have not been observed and to first approximation, we assume that BH-HMXBs of Pop III are not fundamentally different from BH-HMXBs in the local Universe. BH-HMXBs of Pop III result from a top heavy stellar IMF being statistically more massive and forming more frequently than the BHs in the local Universe (Woosley 2017; Di Carlo et al. 2020), with donor stars of primordial chemical composition.

Madau & Fragos (2017) have shown that although the X-ray luminosity in the 2–10 keV range from the HMXB population per unit SFR in galaxies may increase by an order of magnitude from the present to redshift z = 8 (Lehmer et al. 2016), for z > 8 it remains flat as the effects of lower metallicity saturate. If the Madau & Fragos (2017) model is correct, the X-ray and radio wavelength luminosities of BH-HMXBs at CD for a given mass and SFR, could be enhanced at most by a factor of ~10 relative to the local Universe, and the results from the calculations that follow would remain valid at the level of an order of magnitude.

VI.5 Number and total mass in BH-HMXB-MQs required to match the onset of EDGES

As shown before, in the absence of X-ray heating to produce an HI absorption as deep as that reported by EDGES it may be required an additional CRB background with a brightness temperature that is about 1.5 times the brightness temperature of the CMB. We now estimate the total number of discrete BH-HMXB-MQs needed at cosmic dawn to produce such additional background. We emphasize that we are not trying to model the detailed EDGES HI absorption profile including a CXB that may rise the kinetic temperature of the gas, decreasing the depth of the absorption, but only to establish if the minimum required CRB can be produced with population III BH-HMXB-MQs similar to those at the present epoch.

Using the flat radio spectrum (Fender et al. 2000) of the quasi-steady mean flux density at 1.4 GHz of $S_{1.4} = 14.6 \pm 0.5$ mJy and distance $d = 2.23$ kpc (Miller-Jones et al. 2021) of Cygnus X-1, implies a luminosity per frequency interval of

$L_{1.4} = 4\pi d^2 S_{1.4} = 8.7 \times 10^{19} erg s^{-1} Hz^{-1}$.

From $S_\nu(T_{CRB}) = 8.31 \times 10^{-20} erg s^{-1} cm^{-2} Hz^{-1} = 8.31 \times 10^3 Jy$ in eq. (1),

and

$D_L = 224.5$ Gpc $= 6.94 \times 10^{29}$ cm, at $z = 19.5$ in equation (2)

$N = \frac{4\pi D_L^2 S_\nu}{(1+z) L_{1.4}}$ eq. (2)

it is obtained $N = 2.82 \times 10^{20}$, which is a very large number of sources.

VI.6 BH-HMXB-MQs formation rate versus stars formed at CD

Assuming the mass of Cygnus X-1 is 50 $M_\odot$ the total mass of BH-HMXBs as Cygnus X-1 in the redshift interval of $z = 20.5$ to $z = 18.5$ would have to be $1.4 \times 10^{22}$ $M_\odot$. Since the volume and time interval between $z = 20.5$ and $z = 18.5$ are respectively $2.6 \times 10^{11} Mpc^3$ and 28 $Myr$, the BH-HMXB-MQ mass formation rate density in that interval needed to match the onset of the EDGES absorption would have to be

$\sim 1.9 \times 10^4 M_\odot yr^{-1} Mpc^{-3}$.

Models of the Pop III Star Formation Densities (SFRDs) explore different scenarios that result in values for the SFRD at CD that differ by orders of magnitude. For instance, for the Pop III SFRD at z=18-20 have been estimated values of $< 10^{-5} - 10^{-4}$ $M_\odot$ yr$^{-1}$ Mpc$^{-3}$ (e.g. Visbal, Haiman & Bryan 2015; 2018; Mebane, Mirocha & Furlanetto 2018). Much larger SFRDs up to $\sim 3.2 \times 10^{-4}$ $M_\odot$ yr$^{-1}$ Mpc$^{-3}$ are estimated by Jaaks et al. (2019), and Wu et al. (2021) from models with the highest SFR efficiency. Assuming a high efficiency SFRD$\sim 10^{-3.5} M_\odot yr^{-1} Mpc^{-3}$ at z = 18.5-20.5 (Jaaks et al. 2019), a total mass of $\sim 2.3 \times 10^{15} M_\odot$ would have been used to form stars in that interval.

A possible correction to this estimate comes from recognizing that star formation probably started at $z = 30$ instead of at $z = 20.5$. The volume and time interval between $z = 30$ and $z = 20.5$ are $9.3 \times 10^{11} Mpc^3$ and 74 $Myr$, respectively, a factor about 3-4 times larger than between $z = 20.5$ and $z = 18.5$. However, the star formation rate density drops between $z = 20.5$ and $z = 30$, by a factor of $\sim 10^3$ (Jaacks et al. 2019). A numerical integration indicates that the mass going into stars in the $z = 20.5$ to $z = 30$ interval is only ~13% of that in the $z = 20.5$ to $z = 18.5$ interval, which results in a total of ~2.6 x $10^{15}$ $M_\odot$.

Then, comparing the $1.4 \times 10^{22}\ M_\odot$ required in sources like Cygnus X-1 to match the onset of EDGES, with the total stellar mass of $2.6 \times 10^{15}\ M_\odot$ expected to have been formed at the same redshift, there is a discrepancy of a factor of $\sim 5 \times 10^6$.

Therefore, it is concluded that the cosmic radio background required to explain the onset of the EDGES signal cannot be provided by stellar BHs in BH-HMXB-MQs systems like Cygnus X-1 formed at cosmic dawn, unless they are radically different to BH-HMXB-MQs in the local Universe.

VI.7 Highly variable Radio Luminous BH-HMXB-MQs (RLBH-HMXB-MQs)

A possible way to diminish the discrepancy of 6-7 orders of magnitude between the mass of XRBs like Cygnus X-1 and the estimated stellar masses at CD would be to consider that the compact binary sources formed at cosmic dawn are similar to XRBs as Cygnus X-3, SS 433 or GRS 1915+105, that are about $10^2$-$10^3$ times more radio luminous than Cygnus X-1. However, some of those specific types of XRBs would not exist at cosmic dawn. GRS 1915+105 is a BH of $\sim 10\ M_\odot$ fed by Roche lobe overflow from a red giant star of $\sim 1\ M_\odot$. The compact object in SS433 is fed by Roche lobe overflow from a $< 3\ M_\odot$ star transiting the Hertzsprung Gap (King & Begelman 1999). Furthermore, given our assumption of top heavy IMF in Pop III BHs and current models of massive stars evolution, compact objects with masses of $< 3\ M_\odot$ as in SS 433 (2.9 +/- 0.7 $M_\odot$ Hillwig et al. 2004) and Cygnus X-3 (2.1+/- 1.1 $M_\odot$ Zdziarski et al. 2013) are unlikely to have been formed in large numbers during the onset of the EDGES absorption of $\Delta z = 2$, that corresponds to an interval of time of $\Delta t = 28 \times 10^6$ yr, insufficient to accommodate the complete evolution of large numbers of low-mass stars. Anyway, these types of XRBs would diminish the discrepancy at most by two to three orders of magnitude.

Cygnus X-3 is the most radio luminous HMXB in the Galaxy (Trushkin et al. 2017; Egron et al. 2017; 2021 and references therein). Following the same procedure as for Cygnus X-1, from 1550 measurements of the Cygnus X-3 fluxes with GBI it is estimated a mean observed flux of 350 mJy at 21cm. For a distance of ~7 kpc, we infer a radio luminosity about $2 \times 10^2$ times larger than that of Cygnus X-1. Therefore, the contribution of sources like Cygnus X-3 would still be 4-5 orders of magnitude below the CRB needed to match the onset of the EDGES absorption. However, it is not known whether scaled-up versions of Cygnus X-3 with BHs and donor stars of tens of solar masses and orbital periods of a few hours, could be formed in large numbers at the high stellar and gas densities expected at CD.

VI.8 The CXB from BH-HMXBs that produce a CRB that would match EDGES

From our calculation, it is concluded that an unrealistic large number of Pop III HMXB-MQs would be needed to produce the minimum additional radio background required to account for the onset of the EDGES signal. While this type of sources fail to account for the required CRB, we now calculate how much they would contribute to the cosmic X-ray background. This background was identified in the early years of high energy astrophysics (Giacconi et al. 1962) and its origin above 0.3 keV is believed to come from a combination of many unresolved extragalactic sources (Fabian & Barcons 1992).

In Figure 2 (Right) is shown the high-energy spectrum of Cygnus X-1 as measured by the INTEGRAL/IBIS telescope. Cygnus X-1 spends most of its time in the so-called "low" hard state (Jourdain et al. 2014), where its flux density at 20 keV is given by (Rodriguez et al. 2015):

$$E \times F(E) \simeq 5 keV s^{-1} cm^{-2}.$$

At a distance of $d$ =2.23 kpc this implies a luminosity of

$$E \times L(E) \simeq 5 keV s^{-1} cm^{-2} \times 4\pi d^2 = 3.0 \times 10^{45} keV s^{-1}$$

Assuming that there are $N = 2.82 \times 10^{20}$ such sources at z = 19.5, corresponding to a luminosity distance of $D_L = 6.94 \times 10^{29}$ cm, the expected total 1 keV (the original energy of 20 keV redshifted a factor of ~20) flux density per steradian, $F_e$, would be:

$$E \times F_e(E) = \frac{3.0 \times 10^{45} keV s^{-1} \times N}{(4\pi)^2 D_L^2} = 1.1 \times 10^4 keV s^{-1} sr^{-1} cm^{-2}$$

On the other hand, the measured cosmic X-ray background at 1 keV, $F_m$, is given in the same units by (Gilli 2013):

$$E \times F_m(E) \simeq 10 keV s^{-1} sr^{-1} cm^{-2},$$

The observed CXB is three orders of magnitude smaller than that expected for $2.82 \times 10^{20}$ sources similar to Cygnus X-1 at cosmic dawn. We then conclude that this large population of BH-HMXBs would produce an X-ray background far in excess (by ~10³) to what is observed.

It may be argued that the column densities of $N_H \simeq 10^{23} cm^{-2}$ proposed by Ewall-Wice et al. (2018) could absorb significantly the local 20 keV radiation. We look at this possibility in what follows.

At 20 keV the dominant absorption process in pure hydrogen and helium gas is Compton ionization of these atoms, which is more efficient than the photoionization (Yan et al. 1998). In the case of Compton ionization the photon is not absorbed but only scattered, sharing its energy with the ejected electron. At 20 keV the cross section for Compton ionization per hydrogen atom is $\sigma_C(H) = 0.6 \times 10^{-24} cm^2$ while that per helium atom is $\sigma_C(He) = 1.2 \times 10^{-24} cm^2$ (Yan et al. 1998). At lower energies, it is the bound-free photoionization cross section the one that is dominant. At the 13.6 eV threshold for hydrogen photoionization, the cross section is quite large, $\sigma_{ph}(13.6 eV) = 6.3 \times 10^{-18} cm^{-2}$. However, this cross section decreases as $\nu^{-3}$, so that for 20 keV, $\sigma_{ph}(20 keV) = 2.0 \times 10^{-27} cm^{-2}$, much smaller than the Compton cross section.

The total Compton opacity will be $\tau_C(H + He) = N_H \sigma_C(H) + N_{He} \sigma_C(He)$. Assuming one helium atom per every 10 hydrogen atoms the total opacity will be

$$\tau_C(H + He) = N_H \sigma_C(H) + 0.1 N_H \sigma_C(He).$$

For $N_H = 10^{23} cm^{-2}$ we obtain $\tau_C(H + He) = 0.072$,

that implies minor absorption of the 20 keV photons.

It can be argued that these energetic photons are not detected because they are absorbed by neutral hydrogen as they redshift on their way to us, but following the treatment of McQuinn (2012) we find that the Universe is optically thin to $\geq$ 1 keV photons from $z \simeq 20$.

## VII. What type of radio sources match the onset of the HI absorption

In Table 1 we list examples of different observed types of radio sources. In the first column we give the name of the prototypical source and in column 2 its classification. Columns 3, 4 and 5 give the flux density of the source at the frequency of 1.4/(1+z) GHz, its distance (in the case of sources at cosmological distances its luminosity distance), and its luminosity per frequency interval at 1.4 GHz. Column 6 gives the number of these sources required to explain the EDGES absorption calculated with eq. (2), and column 7 and 8 give the characteristic mass of the source and the total mass required to produce these sources (equal to the product of the number of sources times the characteristic mass). Finally, column 9 gives the mass formation rate density of the respective type of source that would be required to match the onset of EDGES, and column 10 lists the references used.

As discussed in section VI.6, the expected maximum mass form rate density in stars between z = 30 and z = 18.5 would correspond to a SFRD $\sim 3.5 \times 10^{-4} M_\odot yr^{-1} Mpc^{-3}$. Checking column (9) of Table 1, we find that the mass FRDs of all types of sources of stellar origin and SMBHs in the local Universe, exceed that SFRD $\sim 3.5 \times 10^{-4} M_\odot yr^{-1} Mpc^{-3}$, with the exception of radio-loud quasars powered by a SMBH such as P172+18. Young SNe as the Crab exceed that stellar mass SFRD by factors of 10-20. As discussed in section VI.1, given the expected top-heavy stellar Initial mass function (IMF), SNe would not exist in large numbers at CD. Furthermore, SNe last only in the order of $\sim 10^4$ yr as strong radio sources, while the sources needed must remain as persistent radio emitters over timescales of 110 Myr.

The radio luminosity per BH mass depends on modes and rates of mass accretion, which in turn depend on the particular properties of the immediate environment of the BH, such as gas volume and/or column density and metallicity.

Therefore, the results listed in Table 1 mean that radio sources of stellar origin and SMBHs with accretion and feedback modes as in the local Universe (e.g. Sgr A*, Cen A, M87), cannot provide the CRB required, and among all types of astrophysical radio sources, only BHs with mass accretion modes and radio emission efficiency as the radio loud SMBH P172+18, can be accommodated within the maximum mass SFRD expected at CD. We are then left with extreme radio-luminous BHs as the only possible viable astrophysical type of sources of a CRB that could match the onset of the EDGES absorption, as proposed by Ewall-Wice et al. (2020) and quoted by Mebane et al. (2020).

## Table 1: Parameters of Radio Sources

| (1) Source | (2) Class | (3) Flux (Jy) | (4) Distance | (5) Luminosity (erg s$^{-1}$ Hz$^{-1}$) | (6) Number of sources | (7) Characteristic Mass ($M_\odot$) | (8) Total Mass ($M_\odot$) | (9) Mass Rate Density ($M_\odot$ Mpc$^{-3}$ yr$^{-1}$) | (10) References |
|---|---|---|---|---|---|---|---|---|---|
| Cyg X-1 | BH-HMXB | 0.015 | 2.2 $kpc$ | $8.7 \times 10^{19}$ | $2.8 \times 10^{20}$ | 50 | $1.4 \times 10^{22}$ | $1.9 \times 10^{3}$ | a,b |
| Cyg X-3 | HMXB | 0.35 | 7.0 $kpc$ | $2.1 \times 10^{22}$ | $1.2 \times 10^{18}$ | 13 | $1.6 \times 10^{19}$ | $2.2 \times 10^{0}$ | c |
| Orion | HII region | 400 | 0.4 $kpc$ | $8.5 \times 10^{22}$ | $2.9 \times 10^{17}$ | $2 \times 10^{3}$ | $5.8 \times 10^{20}$ | $8.0 \times 10^{1}$ | d,e |
| Sgr A* | SMBH | 1 | 8.5 $kpc$ | $8.7 \times 10^{22}$ | $2.9 \times 10^{17}$ | $4 \times 10^{6}$ | $1.2 \times 10^{24}$ | $1.6 \times 10^{5}$ | f,g |
| Crab Nebula | Supernova | 900 | 1.9 $kpc$ | $3.9 \times 10^{24}$ | $6.4 \times 10^{15}$ | 10 | $6.4 \times 10^{16}$ | $8.8 \times 10^{-3}$ | h,i |
| 30 Dor | Giant HII Region | 15 | 49 $kpc$ | $4.3 \times 10^{25}$ | $5.8 \times 10^{14}$ | $5 \times 10^{5}$ | $2.9 \times 10^{20}$ | $4.0 \times 10^{1}$ | j,k |
| M33 | Spiral Galaxy | 5.3 | 765 $kpc$ | $3.7 \times 10^{27}$ | $6.8 \times 10^{12}$ | $8 \times 10^{10}$ | $4.3 \times 10^{23}$ | $5.9 \times 10^{4}$ | l,m |
| Cen A | SMBH | 0.8 | 3.8 $Mpc$ | $1.4 \times 10^{28}$ | $1.8 \times 10^{12}$ | $6 \times 10^{7}$ | $1.1 \times 10^{20}$ | $1.5 \times 10^{1}$ | n,o |
| M82 | Star-forming Galaxy | 8.2 | 3.6 $Mpc$ | $1.3 \times 10^{29}$ | $1.9 \times 10^{11}$ | $5 \times 10^{10}$ | $9.0 \times 10^{22}$ | $1.2 \times 10^{4}$ | p,q |
| M87 | SMBH | 3.0 | 16.4 $Mpc$ | $9.7 \times 10^{29}$ | $2.6 \times 10^{10}$ | $7 \times 10^{9}$ | $2.0 \times 10^{20}$ | $2.7 \times 10^{1}$ | r,s |
| Arp 220 | ULIRG | 0.31 | 77 $Mpc$ | $2.2 \times 10^{30}$ | $1.1 \times 10^{10}$ | $1 \times 10^{11}$ | $1.1 \times 10^{21}$ | $1.5 \times 10^{2}$ | t,u |
| P172+18* | SMBH | 0.006 | 68 $Gpc$ | $4.3 \times 10^{33}$ | $5.8 \times 10^{6}$ | $3 \times 10^{8}$ | $1.7 \times 10^{15}$ | $2.3 \times 10^{-4}$ | v |

*P172+18 is located at z = 6.8 (Bañados et al. 2018). The flux density given is at a frequency of 1.4/(z+1) GHz. The distance is the luminosity distance.

REFERENCES: a) This paper. b) Miller-Jones et al. (2021). c) Zdziarski et al. (2013). d) Felli et al. (1989). e) Kounkel et al. (2017). f) Falcke et al. (1998). g) Reid et al. (2019). h) Vinyaikin (1993). i) Trimble (1968). j) Cersosimo & Loiseau (1984). k) Pietrzynski et al. (2013). l) Terzian & Pankonin (1972). m) Freedman et al. (1991). n) Burns et al. (1983). o) Harris et al. (2010). p) Muxlow et al. (2006). q) Karachentsev & Kashibadze (2006). r) Biretta et al. (1991). s) Tikhonov et al. (2019). t) Ivison et al. (2004). u) Ahn et al. (2012). v) Bañados et al. (2018).

## VII.1 Intermediate Mass BHs (IMBHs) at z ~ 18-20

At present the radio luminosity of BHs scales roughly as their mass to the 1.2 power (Merloni et al. 2003). Then, for a given accretion mass density we may be justified in assuming a linear relation also for CD. The key assumption here is that the radio luminosity to mass ratio in the BHs at CD is similar to that of the most efficient BH radio emitters observed at z = 6-7. Although BHs at CD have not been observed and there is no observational ground for this assumption, the BH seeds at CD of the radio luminous BHs observed at z = 6-7 are expected to beaccreting at high rates, producing copious radiation across the electromagnetic spectrum. A local analog of those BH seeds at CD could be the massive BH at the dynamic center of the dwarf starburst galaxy Henize 2-10 (Reines, Sivakoff1, Johnson et al. 2011).

What 1.4 GHz luminosity per frequency interval can we adopt for instance for $\sim 10^3 M_\odot$ BHs? These BHs would fall in the category of IMBHs, for which at present there are no clearly established observational examples (Greene et al. 2020). We will then assume that the radio luminosity of these IMBHs scales linearly with the mass as in P172+18. We will then use for these IMBHs

$$L_{1.4}(IMBH) = L_{1.4}(P172+18)\frac{10^3}{3\times 10^8} = 1.4\times 10^{28} erg s^{-1} Hz^{-1}. \qquad \text{eq. (3)}$$

Since we are assuming that the radio luminosity scales directly with the mass, the total mass required in these IMBHs is the same as that required in SMBHs as in P172+18-like quasars. This implies that a significant fraction of the mass invested in star formation by $z = 18$ is in IMBHs. An estimate of this fraction can be obtained by dividing the mass of $1.7\times 10^{15} M_\odot$ required by this type of BH sources in Table 1, over the maximum possible total mass expected to be invested in star formation ($2.6\times 10^{15} M_\odot$) to get a mean fraction of 0.7. Therefore, under the hypothesis of a BH CRB, and assuming a BH mass-radio luminosity ratio as in the quasar P172+18, the onset of the EDGES absorption trough would require that at least the equivalent of ~70% of the global BH mass to the expected maximum global stellar mass ratio at cosmic dawn be in BHs ($M_{BH}/M_{stell}$ > 70%).

## VII.2 BH masses versus stellar masses in primordial galaxies

This BH to stellar mass ratio of $M_{BH}/M_{Stell}$ > 70% at CD is very large compared to that in the local Universe. In the Milky Way have been estimated ~1.9 x $10^9$ $M_\odot$ in stellar BHs (Olejak et al. 2020 and references therein), plus a relative tiny contribution by the SMBH of 4.1x$10^6$ $M_\odot$ in the Sgr A* BH (Ghez et al. 2008; Genzel et al. 2010). For an estimated total stellar mass in the Galaxy of 5.5 ± 0.9 x $10^{10}$ $M_\odot$ (McMillan 2011; Licquia et al. 2013), this implies a total BH to total stellar mass ratio $M_{BH}/M_{Stell}$ ~3.5 ± 0.5 % in the Milky Way. Therefore, the >70% global BH to stellar mass ratio in galaxies at CD is >20 times larger than the overall $M_{BH}/M_{Stell}$ ~3.5% ratio estimated in the Milky Way. Considering only the SMBH of 4.1x$10^6$ $M_\odot$ in Sgr A* that mass ratio in the Galaxy would be $M_{BH}/M_{Stell}$ ~ 7.5 x $10^{-5}$, which differs by several orders of magnitude from the global ratio of $M_{BH}/M_{Stell}$ > 0.7 at CD.

We point out that the global $M_{BH}/M_{Stell}$ ratio inferred from our analysis of the onset of the EDGES absorption is consistent with the evolution of the $M_{BH}/M_{bulge}$ ratio in galaxies at high redshifts, where BHs seem to grow faster than their hosts star-forming galaxies. Parameterizing this evolution it is found a redshift evolution in galaxies of $M_{BH}/M_{bulge} \propto (1 +$

z)$^\beta$, with values that span the range of β ≈ 0.7–2 (Kormendy & Ho 2013). These results imply that in the first galaxies the IMBH progenitors of the SMBHs observed at z = 6-7 come first, and are formed by direct collapse, before the rest of the stellar components (e.g. bulges).

VII.3 SMBHs in quasars at z = 5-7

The numbers of massive galaxies and SMBHs in quasars at z > 5 have increased dramatically in recent years and their existence within the first billon years of the Universe remains an intriguing puzzle (Inayoshi,Visbal & Haiman 2020 for a review). Of particular interest for the subject of this review are the radio loud QSOs at high redshifts. Radio jet-enhanced disk accretion in active galactic nuclei (AGNs) are believed to play an important role in the early growth of RLMBHs (Jolley & Kuncic 2008). At z = 5-7 this type of sources can provide useful information to theorize on the formation of the first massive BHs, their galaxy hosts, high density environments and modes of accretion. In addition, RLMBHs may be individual targets for HI absorption studies of the IGM in the EoR and CD.

Out of the ~200 reported quasars at z > 6 only three are presently known to be undergoing a radio-loud phase. P172+18, listed in Table 1, is the most distant currently known radio loud QSO at z=6.823 (Bañados et al. 2021). For a BH mass of ∼3 × 10$^8$ M$_\odot$ and a radio-loudness parameter (R2500 = fv,5 GHz/fv,2500 Å) ∼ 90, P172+18 is proposed as the fastest presently known accreting BH at a cosmic age of ∼800 Myr, close to the end of the EoR. VLBI observations of P172+18 show a resolved compact source of 52.5 × 18.6 pc with flux density of 398.4 ± 61.4 μJy at 1.53 GHz (Momjian et al. 2021). This quasar is not strongly gravitationally lensed because there is no indication of multiple radio components in the field of the source. The derived minimum magnetic field is 0.037 G, and the jet synchrotron losses dominate over inverse Compton scattering off the CMB at z=6.823, with relativistic electron radiative lifetimes of 0.7 yr (Momjian et al. 2021). Since the source size is of ∼50 pc, to maintain that electron distribution, new injection or reacceleration of particles is needed, with an estimated kinematic age of ∼10$^3$ yr. For the following calculations, we use P172+18 as the prototype mass-growing RLMBH at the end of the EoR.

Another remarkable RLMBH is PSO J352.4034–15.3373, that at z = 5.84 has a radio luminosity about 10 times larger than that of P172+18 (Bañados et al. 2018). Unfortunately, the mass of the associated BH in this quasar is not known yet and we cannot use this source for our comparisons. But if the mass were similar to that of P172+18 (3 × 10$^8$ M$_\odot$), the BH mass requirements to provide the CRB would become lower by an order of magnitude. It is hoped that the determination of the BH masses in these high z quasars will be improved by observations with the JWST.

It is interesting that the SMBH mass to total stellar mass (M$_{BH}$/M$_{Stellar}$) ratio in these quasars at z = 6-7 with SMBHs of > 10$^8$ M$_\odot$ seems to be higher by factors 3 – 4 than expected from local relations, which suggests that SMBHs at high redshifts grow more rapidly than the host galaxies (Rojas-Ruiz, Bañados, Neeleman et al. 2021). In section VII.5b we have shown that the onset of the EDGES absorption at z =18 - 20, would suggest that in primordial galaxies the ratio of SMBH masses to host galaxy total stellar mass could be even higher, by factors of ~20, than expected from local relations.

VII.4 BH seeds of SMBHs at z = 5-7

To explain the origin of these SMBHs at z = 6-7 several models based on astrophysical and non-astrophysical types of BH seeds have been theoretically proposed (for a comprehensive review see Inayoshi et al. 2020). Presently, there is no concluding evidence that those SMBHs are formed via one or several of the many proposed scenarios. In this context, we briefly refer here to some of the BH-seed evolutionary scenarios for the formation of SMBHs of ~$10^9$ $M_\odot$ at z = 6-7, that may be consistent with IMBHs of $10^{3-5}$ $M_\odot$ at cosmic dawn (z~20) as the sources of a CRB that matches the onset of EDGES.

   a) IMBH growth at CD from typical Pop III BH remnants of stellar seeds of 10-$10^2$ $M_\odot$

Given the large column densities of gas that should prevail in the early universe, it may be expected that a very rapid growth of typical Pop III stellar BHs to SMBHs may take place by mean Eddington accretion rates with episodes of super-Eddington rates, (Begelman 1979; Abramowicz, Czerny, Lasota et al. 1988; Volonteri, Silk & Dubus 2015; Eishun et al. 2019; Hardcastle & Croston 2020).

The growth of BH mass under Eddington accretion is given by:

$$M_{BH}(t) = M_s exp(t/\tau_{Edd})$$

where $M_{BH}(t)$ is the mass of the BH at time $t$, $M_s$ is the initial "seed" mass of the BH at $t = 0$ and $\tau_{Edd}$ is the Eddington timescale, usually taken to be 45 Myr.

One possible scenario is to simply assume growth at a mean Eddington rate with the occurrence of relatively short super-Eddington accretion phases, initiated in chaotic episodes of uncorrelated initial directions on BHs as in low spin models (e.g. Zubovas & King 2021), that may be followed by radio jet-enhanced disk accretion (Jolley & Kuncic 2008). These episodes of super-Eddington accretion may be possible if at those early times seed BHs of 30-70 $M_\odot$ are formed by direct collapse with no energetic SNe (Heger et al. 2003), and the local column densities of the accreting gas are very large, as suggested by observations of SMBHs in quasars at z > 6 (Venemans et al. 2020). Under those early conditions, one can start with a BH seed of ~ $85 M_\odot$, the largest stellar BH mass so far detected in BHs fusion by LIGO/Virgo (Abbott et al. 2020) that is slightly larger than the ~70 $M_\odot$ lower boundary for the upper Pop III BH gap (Woosley 2017; Woosley & Heger 2021), and reach intermediate mass of ~ $10^3 M_\odot$ by $z = 18$ and of ~ $3\times10^8 M_\odot$ by $z = 7$ as in P172+18. This last mass agrees with the average of the estimated masses of the most massive quasars observed at z > 6 (Inayoshi et al. 2020).

It is relevant to emphasize that if the luminosity per frequency interval at 1.4 GHz scales at CD linearly with the mass of the BH, we could get the same CRB with smaller (larger) BHs in a correspondingly larger (smaller) numbers. It is most likely that SMBHs of ~$10^9$ $M_\odot$ in quasars at z=6-7 represent the tip of an iceberg, but it is not known whether BHs in the early universe fill a continuum of masses (Inayoshi et al. 2020), for instance, from stellar BHs of ~30 $M_\odot$ to SMBHs of ~$10^9$ $M_\odot$, reflecting a range of initial seed masses and different modes of BH accretion.

The vast majority of Pop III stars are expected to be formed in ~$10^{5-6} M_\odot$ dark matter (DM) minihalos with primordial gas undergoing molecular hydrogen ($H_2$) cooling. Only a tiny minority of early BHs, born in highly biased regions of the Universe, grow to ~ $10^{8-9} M_\odot$ by z ~6, the vast majority of BHs being born in the most typical regions, remaining far below those high masses by z~6 (Inayoshi et al. 2020).

From analogous arguments Ewall-Wice et al. (2020) propose two models that are in approximate agreement with EDGES, matching the location and steepness of the HI absorption trough. The parameters of these models correspond to BH formation in minihalos and large halos of virial temperatures of $10^{3-4}$ K and $1-5 \times 10^4$ K, respectively, with BH seed masses of $10^2$ $M_\odot$ and $1.5 \times 10^3$ $M_\odot$ respectively, and super-Eddington accretion at rates of ~1.8 and ~2.5 Eddington. Assuming a mean semi-continuous accretion at Eddington rate and an initial mass of $85 M_\odot$ at z~30, by our calculation we obtained a slightly smaller BH mass of ~ $10^3 M_\odot$ by z~18 and $3 \times 10^8 M_\odot$ by $z = 7$, as in the quasar P172+18.

One possible problem to sustain a mean semi-continuous accretion at Eddington rate may be the limitation imposed by the mechanical feedback observed in accreting BHs of all mass scales. The observation of energetic shocks at large-scale distances from the BH engines as in Cygnus X-1 (Fig. 2), LMC X-1 (Hyde et al. 2017), and compact stellar ULX sources (Feng & Soria 2011; Soria et al. 2021), provide evidence that the power output of the overwhelming majority of stellar BHs is dominated by the kinetic energy of radio 'dark' outflows, whose key signature is the eventual energization of the ambient medium. Energetic mechanical feedback from accreting SMBHs is also observed in galaxies, groups and clusters (McNamara & Nulsen 2012). Mechanical feedback from AGN type radio jets may also have been observed in dwarf galaxies up to redshift z~3.4 (Mezcua et al. 2019). Despite these observational evidences, the question on how significant feedback in the form of kinematic outflows may be to limit the growth of stellar mass BHs at CD remains open.

In addition, several publications have highlighted the difficulty of making IMBHs ($10^3 - 10^5$ $M_\odot$) from merging typical Pop III BH remnants, due to the long dynamical friction timescales (Ma et al. 2021), or growing them via accretion (Smith et al. 2018).

b) SMBHs from DCBHs of very massive gas clouds and stars

Another alternative is the formation of IMBHs in the central regions of proto-galaxies by direct collapse of massive gas clouds, and/or the collapse of supermassive stars formed out of those massive gas clouds, without the need for 'seed' black holes left over from early typical Pop III star formation. It has been shown that BH massive seeds at high redshift could be formed in Atomic Cooling Halos (ACHs) by monolithic collapse of massive gas clouds with virial temperatures Tvir ~$10^4$ K and low angular momentum, because in this way it is avoided gas fragmentation and the formation of dense clumps with subsequent typical Pop III star formation (e.g. Begelman, Volonteri & Rees 2006; Volonteri 2012; Visbal, Haiman & Brayan 2014; Wise et al. 2019; Inayoshi et al. 2020; Lupi, Haiman & Volonteri 2021). The avoidance of gas fragmentation in dense clumps may be due to Lyman–Werner feedback from massive stars (Visbal et al. 2014), or to the dynamics of structure formation by rapid gas inflow onto an unstable central core (Wise et al. 2019).

Supermassive stars could also be formed out of those massive gas clouds that end as massive BHs by direct collapse (Shapiro & Shibata 2002; Bromm and Loeb 2003). Even if those requirements are not achieved for the formation of supermassive BHs of $10^6$ M$_\odot$, relatively lower-mass but still massive seeds could be formed in the IMBH range ($10^3 - 10^5$ M$_\odot$). These IMBH seeds should be abundant and have a significant contribution to the overall mass density of the BH population at z > 18. This population of massive BHs formed by direct collapse (DCBHs) may be within reach of the JWT (Wise et. al. 2019).

c) Primordial Black holes (PBHs) as seeds for the formation of SMBHs

More than 50 years ago, it was proposed that PBHs might result from the collapse of over densities in the very early Universe (Zel'dovich and Novikov 1967; Hawking 1971). PBHs have been conjectured as potential seeds for the formation of SMBHs in high z quasars (e.g. Bean and Magueijo 2002) and as the sources that may boost the EDGES HI absorption (e.g. Hektor, A. et al. 2018). Besides, PBHs have been proposed to be the LIGO-Virgo BHs of 30-40 $M_\odot$ (e.g. Bird S. et al.2016), and potential contributors to the dark matter content of the Universe (e.g. Chapline 1975; Hasinger 2020; and review by Carr et al. 2021).

VII.5 IMBHs at cosmic dawn and their dearth in the local Universe

Although IMBHs could have been formed in large numbers at CD there is some relative dearth of them in the local Universe. There are some important observational candidates for IMBHs, but there have been no concrete observational evidence of BHs with masses of $10^3 - 10^5$ $M_\odot$ in the local Universe (Green, Strader & Ho 2020). Besides, no IMBH has been detected in the Milky Way Galaxy, despite the estimated large numbers of stellar mass BHs. These facts may suggest that IMBHs have difficulties to be formed from stellar BHs, and most may be formed by direct collapse and necessarily grow to SMBHs, under the very high densities of gas that allow their formation at CD and rapid growth to SMBHs in the EoR.

## VIII. IMBHs in the early Universe grow embedded in large gas column densities

SMBHs at high redshifts are observed to be embedded in very large column densities of gas. The most luminous galaxies observed at z = 2–4.6 with the Wide-field Infrared Survey Explorer (WISE) revealed at z > 3 infrared warm hyper-luminous galaxies with bolometric luminosities up to $L_{bol} \sim 10^{14} L_\odot$, and column densities of $N_H = 2–3 \times 10^{24}$ atoms cm$^{-2}$ in the line of sight to the deeply obscured central sources (Tsai et al. 2015; Vito et al. 2018).
On the other hand, observations of quasars at redshifts of z = 6.0-7.5 with the Atacama Large Millimeter/Submillimeter Array (ALMA) show that the majority of these luminous SMBHs in quasars reside in galaxies with molecular gas mass fractions of more than 10% of the total baryonic mass, and star formation rate densities in the central regions (SFRDs) of $10^2$-$10^3$ $M_\odot$yr$^{-1}$ kpc$^{-2}$ (Venemans et al. 2017 & 2020 for a large ALMA survey of QSOs at z~6; Neeleman et al. 2021; Wang et al. 2021 and Yang et al. 2020 for QSOs at z~7.5).

Venemans et al. (2020) point out that the general properties of these ultraluminous quasar host galaxies at z = 6.0-7.5 have some analogous properties to the very rare Ultraluminous Infrared Galaxies (ULIRGs) in the local Universe (e.g. review by Sanders & Mirabel 1996). All ULIRGs up to z~0.1 are advanced mergers of gas-rich galaxies with nuclear regions enshrouded in large column densities of gas and dust. For instance, in Arp 220, the nearest ULIRG listed in Table 1, a column density of molecular hydrogen in excess of $\sim 10^{26} atoms\ cm^{-2}$ is found along the line of sight to the nuclear region (Scoville et al. 2017).

As has been pointed out (e.g. Haiman 2011; Mesinger et al. 2013) to explain EDGES it is required that the radio luminous BHs do not heat the gas and increase the spin temperature of the HI, diminishing the depth of the HI absorption. One possible way out of this has been to assume that BHs in the early universe grow in very high gas density environments with mean column densities $N_H > 10^{23}$ cm$^{-2}$ that absorb the soft X-rays (Ewall-Wice et al. 2018; 2020). This is a reasonable hypothesis since massive BHs in high z quasars are observed to be embedded in even larger column densities of gas.

# IX. Observation of IMBHs at cosmic dawn in radio, infrared and X-rays

IX.1 Observation of IMBHs at cosmic dawn in the radio continuum and HI absorption

It has been pointed out that the radio sources of the CRB required by EDGES, such as BHs (Ewall-Wice et al. 2020) and galaxies (Reis et al. 2020), could be detected individually or as a significant modified shape and time evolution of radio emission fluctuations in ultrasensitive surveys at long radio wavelengths with interferometers, which have the advantage of suppressing the extended foreground emission that limits all-sky detectors as EDGES. Following our calculations, such modified shape and time evolution should be dominated by the radiation from growing massive BHs in protogroups and protoclusters of galaxies.

We now estimate if SKA will be able to detect the BHs existing at the beginning of the EDGES absorption at $z = 20.5$ as individual continuum sources, and if HI could be detected in absorption against the continuum emission from those sources. We assume that the radio luminosity of these BHs scales linearly with the mass taking as reference P172+18. We then have that their radio luminosity at 1.4 GHz will be given by:

$$L_{1.4}(BH) = L_{1.4}(P172+18)\frac{M_{BH}}{3\times10^8} = 1.4\times10^{25}M_{BH}\,erg\,s^{-1}Hz^{-1},$$

where $M_{BH}$ is the mass of the BH in solar masses. The flux density observed in the present Universe at a frequency of $1.4 GHz/(z+1) = 65 MHz$ will be:

$$S_{65MHz}(BH) = (1+z)\frac{L_{1.4}(BH)}{4\pi D_L^2} = 4.5\times10^{-6}M_{BH}\,\mu Jy,$$

where $D_L = 237 Gpc$ is the luminosity distance. In the vicinity of 100 MHz, SKA1-low will have a continuum rms of 0.8 $\mu Jy$ for a 1000-hour integration (Mellema et al. 2015). If sources with a mass equal to that of P172+18 existed at that redshift, they would be easily detectable with a flux density of $S_{65MHz} = 1.4$ mJy. More realistically, at that epoch we expect that the most massive black holes would have masses in the order of $10^6 M_\odot$, presenting a flux density of a few $\mu Jy$ at 65 MHz. These sources would be detected at marginal levels.

The detection of HI in absorption will be even more difficult. The line rms of SKA1-low with a frequency resolution of 1 MHz is about 5 $\mu Jy$ for a 1000-hour integration. The significative detection of the absorption feature (that is expected to have levels comparable to the continuum) will only be possible for BHs with masses in excess of $10^7 M_\odot$. Alternatively, a protogroup with tens of galaxies each with radio-loud central BHs would provide a larger signal. Furthermore, some of these remote BHs could have a blazar geometry and be detected at levels one to two orders of magnitude larger that their isotropic luminosity, or be amplified by intervening gravitational lenses. Surveying large areas of the sky should lead to the detection of some of these sources and to further progress on studies of the HI absorption at cosmic dawn.

IX.2 Observations of IMBHs at cosmic dawn in infrared wavelengths

JWST will make possible to investigate at infrared wavelengths the formation of structures up to z = 8-10, which would provide evidence for the formation of stars and BHs out to z ~ 15 (Robertson 2022). The galaxy hosts of massive DCBHs are expected to have absolute magnitudes MAB < 25 that should be unambiguously detectable by the Mid-Infrared Instrument (MIRI), whereas galaxies with growing Pop III stellar remnants with predicted absolute magnitude MAB > 30 will likely be undetectable (Natarajan, Pacucci, Ferrara et al. 2017). These authors propose that MIRI observations may be the way to discriminate between the initial BH seeding mechanisms for the formation of SMBHs at z > 6 described in sections VII.4 a) & b). In the early transitory phase of the highest z galaxies hosts of DCBHs accretion onto the DCBH is expected to outshine the sub-dominant stellar population. This results in distinct, detectable spectral signatures in the infrared (Natarajan, Pacucci, Ferrara et al. 2017). Contrary to the local $M_{BH}/M_{buldge}$ of up to ~0.5% in galaxies of the local universe, in this transient early phase the galaxy hosts of DCBHs would be characterized by $M_{BH}/M_{buldge} > 1$ (Agarwal et al. 2013; Kormendy & Ho 2013).

IX.3 Observations of IMBHs at cosmic dawn in X-rays

At present only a few AGNs at z > 6 have been firmly detected in the X-rays (Bañados et al. 2018; Pons et al. 2020; Connor et al. 2020). Future X-ray missions such as Athena, AXIS and LynX will certainly push the detection limits to earlier epochs. However, the predicted number density of AGNs detectable in the X-rays span at least an order of magnitude and make predictions uncertain (Habouzit et al. 2022).

# X. Conclusions

1) Recent observations of SMBHs of ~$10^9$ $M_\odot$ in radio loud quasars at redshifts z = 5 to 7 (Bañados et al. 2018; 2021; Momjian et al. 2021) suggest the existence of a CRB produced by earlier progenitors of those SMBHs. By simple calculations we have shown that among all types of astrophysical radio sources, rapidly growing IMBHs (Ewall-Wice et al. 2018, 2020) are the dominant radio sources in primordial galaxies with extreme radio luminosity to star formation rate (Mirocha et al. 2018; Mirocha & Furlanetto 2019; Mebane et al. 2020), that can produce at cosmic dawn (z = 18-20) a CRB that boosts the amplitude of the redshifted 21cm HI absorption (Table 1 and discussions in section VII).

2) Assuming a BH mass to radio luminosity ratio as observed in radio-loud quasars at z > 6, the large amplitude of the onset of HI absorption in the interval from z = 20.5 to 18.5 (~28 Myr) reported by EDGES (Bowman et al. 2018a), can be matched by a CRB from rapidly growing radio luminous IMBHs accreting mass at very high rates. That CRB implies that the global mass density of IMBHs at cosmic dawn must be dominant over that of stars, more than 70% of the maximum estimated stellar mass density at those redshifts (section VII.1).

3) The global ratio of $M_{BH}/M_{Stell} > 0.7$ at cosmic dawn is at least 20 times larger than the estimated total $M_{BH}/M_{Stell} \sim 4\%$ ratio in the Milky Way galaxy. This transition from a global $M_{BH}/M_{Stell} > 70\%$ ratio to the $M_{BH}/M_{Stell} \sim 4\%$ is consistent with the observed evolution of that ratio in galaxies up to high redshifts. This means that the IMBH progenitors of the SMBHs observed at z = 6-7 are formed first, before the bulk of stars, with no need of mass contribution from stellar BH remnants of typical Pop III stars (section VII.2).

4) The onset of the EDGES absorption between z = 18.5 to 20.5 cannot be explained by BH X-ray binaries as observed today (e.g. Cygnus X-1, Cygnus X-3). The unreasonable large number of sources of this type that are required to match the onset of the EDGES trough would exceed the expected maximum total mass of stars already formed at that epoch by factors of $\sim 10^6$ for sources like Cygnus X-1, and factors of $\sim 10^4$ for sources like Cygnus X-3. Furthermore, the cosmic X-ray background (CXB) component produced by the required numbers of BH-XRBs like Cygnus X-1 and Cygnus X-3 to match the onset of the EDGES absorption, would exceed respectively, by factors of $10^3$ and 10 the observed CXB. (sections VI.5, VI.6 & VI.7).

5. We show that the inefficiency of BH-HMXB-MQs like Cygnus X-1 to produce radio emission at 21cm is due to inverse Compton (IC) cooling by the photons from the donor star, which affects the microscopic motions of the radiating relativistic electrons, suppressing the synchrotron emission, but not affecting the bulk motion of the ejecta. This is why these BH-XRBs produce energetic mechanical feedbacks with radio "dark" jets (section VI.3).

6. SMBHs in quasars at z > 6 are observed to be embedded in local gas column densities of $N_H \sim 10^{23-24}$ atoms cm$^{-2}$. These large column densities of gas block the potentially heating X-rays of <1 keV produced by the same BHs, but are transparent to the 21cm radio continuum radiation (section VIII). IMBHs at z = 18 - 20 are likely to be embedded in similar or even larger column densities of neutral gas. Those extreme column densities of neutral gas at cosmic dawn could be the condition for the formation of IMBHs, their rapid growth, and final end as SMBHs at z ~ 7. In this context, the majority of IMBHs would be transient objects, mostly existing at cosmic dawn and EoR, which could explain their dearth relative to stellar mass BHs in the local Universe (section VII.5).

7. In section IV several possible obstacles and caveats to a synchrotron CRB of astrophysical origin have been analyzed and discussed:

a) Inverse Compton photons of the CMB at cosmic dawn cannot suppress a synchrotron CRB from the magnetized compact jets of BH-XRBs and SMBHs. We have shown that at z~20 the synchrotron cooling time in BH-jets dominate respectively, by factors of more than $10^4$ and up to $10^9$ the Inverse Compton time of the CMB. The radio emission of supernova remnants (SNRs) have magnetic fields of only a few μGauss and is suppress by the CMB at cosmic dawn. Radio emission in supernova explosions (SNe) have larger magnetic field intensities but last only few years, while the CRB needed to match EDGES would require that they last tens of millions of years (section IV.1).

b) X-rays from the same BHs that produce the radio emission cannot heat the HI gas on large distance scales. IMBHs at cosmic dawn accrete mass at high rates from local gas column densities of $N_H > 10^{23}$ atoms cm$^{-2}$ that block the potentially heating soft X-rays of < 1 keV (section IV.2).

c) The combined heating impact on the HI due to subtle effects of the CMB and CRB from BH-jets at cosmic dawn (z = 18-20) is less than 10% of the gas temperature at that epoch. Therefore, these subtle, second order effects have been ignored in our approximate calculation of the brightness temperature of the BH-CRB at cosmic dawn (section IV.3).

8. The extended foreground emission that limits all-sky detectors as EDGES will be resolved out in observations with interferometers such as SKA. Long (1000-hour) integrations with this instrument will be able to detect the radio luminous BH sources responsible for HI absorptions such as that reported by EDGES, as individual sources in the continuum. HI line absorption will be detectable in the case of proto-groups of galaxies with central growing BHs, and even individually in the case of being cosmic dawn blazars, or amplified by intervening gravitational lens sources (section IX).

9. There are three possible outcomes on the EDGES signal at Cosmic Dawn:

   i)   The onset of the EDGES signal is confirmed by independent observations. In this case, the conclusions 1, 2, and 3 on the IMBHs at CD and EoR will stimulate the exploration of these objects at radio wavelengths with ALMA, ngVLA and SKA, in the infrared with the JWST, and with future new generation telescopes for X-ray observations.

   ii)  Future measurements may show that the absorption in the EDGES signal is not of cosmic origin and due to systematics, signal extraction biases, etc.

   iii) Future observations find an absorption signal of reduced or no additional amplitude. Here we provide simple approximate calculations that other observers can use to extract information on the BHs formed in the early Universe.

In the alternatives i and iii, the additional amplitude of the HI absorption at z = 18-20 would be an indirect signal of a population of rapidly growing BHs that can explain the existence of SMBHs of ~$10^9$ M$_\odot$ in QSOs at z = 6 up to z ~ 7.5 (e.g. Bañados et al. 2018; Yang et al. 2020, Wang et al. 2020b; Wang et al. 2021), grown at high accretion rates, in rare dark matter halos, when the Universe was less than 150 Myr old (e.g. Bromm & Loeb 2003; Begelman, Volonteri & Rees 2006; Kaurov et al. 2018; Fialkov & Barkana 2019; Inayoshi, Visbal & Haiman 2020). The most radio loud BHs, blazar BH sources and gravitational lens amplified BH radio sources at cosmic dawn, could be detectable at the $\mu Jy$ level with SKA, and used as targets for HI absorption studies of the IGM at Cosmic Dawn and Epoch of Reionization.

Acknowledgements: We thank the editor James Miller-Jones and an anonymous referee for their comments that produced significant improvements of the originally submitted manuscript. I.F.M. thanks Philippe Laurent for his input on the high-energy observations of Cygnus X-1 with the INTEGRAL satellite, and David Elbaz, Dave Sanders, Ed van den Heuvel, Norbert Langer, Andrew King, Piero Madau, Jose Groh, Jerome Orosz, Tsvi Piran, Michela Mapelli, Segei Trushkin, and Julien Girard, for their kind responses to specific questions. L.F.R. acknowledges the support of DGAPA, UNAM and CONACyT, México.

References

Abbott, R., Abbott, T. D., Abraham, S., et al. 2020, ApJ, 900, L13
Abramowicz, M. A., Czerny, B., Lasota, J. P. et al. 1988 ApJ 332, 646

Ahn, C. P., Alexandroff, R., Allende Prieto, C., et al. 2012, ApJS, 203, 21
Agarwal, B., Dalla Vecchia, C., Johnson, J. L., et al. 2014, MNRAS, 443, 648
Anhanna, T.T. et al. 2019 ApJ 871, 240
Allan, A.P., Groh, J. H., Mehner, A. et al. 2020 MNRAS 496,1902-1908
Baczko, A.-K., Schulz, R.; Kadler, M. et al. 2016 A&A 593, A47
Bañados, E., Connor, T., Stern, D. et al. 2018 ApJ L 856, L25
Bañados, E., Venemans, B. P., Mazzucchelli, C., et al. 2018, Nature, 553, 473
Bañados, E., Mazzucchelli, C. & Momjian, E. 2021 ApJ 909, 80
Barkana, R. 2018, Nature 555, 71-74
Basu-Zych, A. R.; Lehmer, B. D.; Hornschemeier, A. E., et al. 2013, ApJ, 774, 152B
Bean R. and Magueijo J. 2002 Phys. Rev. D 66 063505
Begelman M. C. 1979. MNRAS 187:237–51
Begelman, M. C., Volonteri, M. & Rees, M. J. 2006, MNRAS 370, 289-298
Bestenlehner, J. M., Crowther, P. A., Caballero-Nieves, S. M. et al. 2020, MNRAS, 499, 1918–1936
Bevins, H. T. J et al. 2021MNRAS 508, 2923B
Bird S. et al. 2016 Phys. Rev. Lett. 116 201301
Biretta, J. A., Stern, C. P. & Harris, D. E. 1991, AJ, 101, 1632
Bolton, C. T. 1972, Nature 235, 271-273
Bowman J. D. et al., 2013, PASA, 30, e031
Bowman, J. D., Rogers, A. E E., Monsalve, R. A. et al. 2018a, Nature, 555, 67
Bowman, J. D., Rogers, A. E E., Monsalve, R. A. et al. 2018b, Nature, 564, E35
Bowyer, S., Byram, E. T., Chubb, T. A. et al. 1965, Science, 147 (3656): 394–398
Braes, L. L. E. & Miley, G. K. 1971 Nature 232, 246B
Brandenberger R., Cyr B., Schaeffer T., 2019, J. Cosmol. Astroparticle Phys., 2019, 020
Brorby, M., Kaaret, P., Prestwich, A. et al. 2016, MNRAS, 457, 21
Bromm, V., Coppi, P. S. & Larson, R. B. 2002, ApJ, 564, 23
Bromm V. & Loeb A., 2003, ApJ, 596, 34
Bromm V. & Yoshida N. 2011. Annu. Rev. Astron. Astrophys. 49:373–407
Burns, J. O., Feigelson, E. D. & Schreier, E. J. 1983, ApJ, 273, 128
Carr, B. et al. 2021 Rep. Prog. Phys. 84, 116902
Cersosimo, J. C. & Loiseau, N. 1984, A&A, 133, 93
Chapline G. F. 1975 Nature 253, 251
Chatterjee, A, Dayal, P., Choudhury, T. R. et al. 2020, MNRAS 496, 1445–1452
Cherepashchuk, A., Postnov, K., Molkov, S., et al. 2020, NewAR, 89, 101542
Chianese M., Di Bari P., Farrag K. et al. 2018 (arXiv:1805.11717)
Condon, J. J., Cotton, W., Fomalont, E., et al. 2012, ApJ 758, 23
Condon, J. J. & Ransom, S. M. 2016, Essential Radio Astronomy. ISBN: 978-0-691-13779-7 Princeton, NJ: Princeton University Press.
Connor, T., Stern, D., Bañados, E. et al. 2021 ApJL 922, L24
Chuzhoy L. & Shapiro P. R., 2007, ApJ, 655, 843
DeBoer D. R. et al., 2017, PASP 129, 045001
Dhawan, V., Mirabel, I. F., Rodríguez, L. F. 2000, ApJ 543, 373D
Di Carlo, U. N., Mapelli, M., Giacobbo, N. et al. 2020 MNRAS 498, 495-506
Douna, V. M., Pellizza L. J., Mirabel, I. F. et al. 2015, A&A 579, A44
Douna, V. M. et al. 2018 MNRAS 474, 3488-3499
Dowel, J. & Taylor, G.B. 2018, ApJL, 858, L9
Eishun, T., Inayoshi, K., Ohsuga, K. et al. 2019 MNRAS 488, 2689–2700
Egron, E. et al. 2017, MNRAS 471, 2703–2714
Egron, E. et al. 2021, ApJ 906, 10 (9pp)

Ellingson S. W. et al. 2009, IEEE Proc., 97, 1421
Ewall-Wice, A., Chang, T.-C, Lazio, J. et al. 2018, ApJ 868, 63
Ewall-Wice, A., Chang, T.-C & Lazio, J. 2020, MNRAS 492, 6086
Fabian, A.C. & Barcons, X. 1992, ARAA, 30, 429
Fabrika S., 2004, Astrophys. Space Phys. Rev., 12, 1
Fabrika, S., 2017 ASPC 510, 395
Falcke, H., Goss, W. M., Matsuo, H., et al. 1998, ApJ, 499, 731
Falcke, H., Körding, E., & Markoff, S. 2004, A&A, 414, 895
Felli, M., Churchwell, E., & Wood, D. O. S. 1989, IAU Colloq.120: Structure and Dynamics of the Interstellar Medium, 315
Fender, R. P., Pooley, G. G., Durouchoux, P., et al. 2000, MNRAS, 312, 853.
Feng, C. & Holder, G. 2018 ApJL, 858, L17
Feng, H. & Soria, R. 2011, New Astronomy Reviews 55, 166–183
Fialkov, A. & Barkana, R. 2019 MNRAS 486, 1763-1773
Fialkov, A., Barkana, R. & Cohen, A. 2018 PhRvL.121, 011101
Field G. B., 1958, Proc. IRE, 46, 240
Finkelstein, S. L., D'Aloisio, A., Paardekooper, J-P. et al. 2019, A.J. 879, 36
Fixsen D. J., Kogut A., Levin, S. et al. 2011, ApJ, 734, 5
Fragos, T., Lehmer, B. D., Naoz, S. et al. 2013, ApJ, 776, L31
Fraser S. et al. 2018, Phys. Lett. B, 785, 159
Freedman, W. L., Wilson, C. D. & Madore, B.~F. 1991, ApJ, 372, 455
Furlanetto, S. R. 2006, MNRAS 371, 867
Furlanetto, S. R., Oh, S. P., & Briggs, F. H. 2006b, Phys. Rep., 433, 181
Gallo, E., Degenaar, N. & van den Eijnden, J. 2018 MNRAS 478, 132
Gallo, E., Fender, R., Kaiser, Ch. et al. 2005, Nature, 436, 819-821
Genzel, R., Eisenhauer, F. & Gillessen, S. 2010 Rev. Mod. Phys. 82, 3121
Ghez, A. M., Salim, S. Weinberg, N. N. et al. 2008 ApJ. 689, 1044-1062
Ghisellini, G., Celotti, A., Tavecchio, A. et al. 2014 MNRAS 438, 2694–2700
Giacconi, R., Gursky, H., Paolini, F.R., et al. 1962, PhysRevLett.9.4399, 439.
Gilfanov M., Grimm H.-J., Sunyaev R., 2004, MNRAS, 347, L57
Gilli, R. 2013 MmSAI 84, 647
Glover S., 2013, in Wiklind T., Mobasher B., Bromm V., eds, Astrophysics and Space Science Library, Vol. 396, The First Galaxies. SpringerVerlag, Berlin, p. 103
Glover, S. & Brand, P. 2003 MNRAS 340, 210–226
Gregory, C. Meynet, G.,Walder, R., et al. 2009, A&A, 502, 611
Greene, J. E., Strader, J. & Ho, L. C. 2020 ARAA 58, 257-312
Gregory, C. Meynet, G.,Walder, R., et al. 2009, A&A, 502, 611
Habouzit, M., Somerville, R. S., Li, Y., et al. 2022, MNRAS 509, 3015
Haiman, Z. 2011, Nature 472, 47
Hardcastle, M. J., & Croston, J. H. 2020, NewAR, 88, 101539
Harris, G. L. H., Rejkuba, M. & Harris, W. E. 2010, Publications of the Astronomical Society of Australia, 27, 457
Hasinger, G. 2020, JCAP, 2020, 022
Hawking, S. 1971 MNRAS 152:75-78
Heger, A. & Woosley, S. E. 2002, ApJ, 567, 532-543
Heger, A. Fryer, C. L.,Woosley, S. E., et al. 2003, ApJ, 591, 288
Heger, A. & Woosley, S. E. 2010, ApJ, 724, 341
Hektor, A. et al. 2018 Phys. Rev. 98, (023503)
Hillwig, T. C., Gies, D. R., Huang, W. et al. 2004 ApJ 615, 422–431
Hills, R., Kulkarni, G., Meerburg, D. et al. 2018, Nature, 564, E32-E34

Hogg, D. W., 1999, arXiv:Astro-ph/9905116v4
Hyde, E. A., Russell, D. M. & Ritter, A. 2017 PASP 129, 094201
Inayoshi, K., Visbal, E. & Haiman, Z. 2020, ARAA 58: 27–97
Ivison, R. J., Greve, T. R., Serjeant, S., et al. 2004, ApJS, 154, 124
Jaacks, J., Finkelstein, S. L. & Bromm, V. 2019 MNRAS 488, 2202–2221
Jana, R., Nath, B.B. & Biermann, P. L. 2019, MNRAS 483, 5329–5333
Jeon, M., Pawlik, A. H., Bromm, V. et al. 2014, MNRAS 440, 3778–3796
Jolley, E. J. D., & Kuncic, Z. 2008, MNRAS, 386, 989
Jourdain, E., Roques, J. P. Chauvin, M. et al. 2012, ApJ 761, 27
Jourdain, E., Roques, J. P. & Chauvin, M. 2014, ApJ 789, 26
Kaaret, P. 2014, MNRAS, 440, L26
Karachentsev, I. D. & Kashibadze, O. G. 2006, Astrophysics, 49, 3
Kaurov, A. A, Venumadhav, T., Dai, L. & Zaldarriaga, M. 2018, ApJL 864, L15
King, A. R. & Begelman, M. 1999, ApJ 519: L169–L171
Kochanek, C. S., Beacom, J. F., Kistler, M. D. et al. 2008 ApJ 684, 1336 .
Koljonen, K. I. I. & Tomsick, J. A.2020, A&A 639, A13
Kormendy, J., & Ho, L. C. 2013, ARA&A, 51, 511
Kounkel, M., Hartmann, L., Loinard, L., et al. 2017, ApJ, 834, 142
Laurent, P., Rodriguez, J., Wilms, J. et al. 2011, Science 332, 438L
Lehmer, B. D.; Basu-Zych, A. R.; Mineo, S. et al. 2016, ApJ 825, 7L
Licquia, T. & Newman, J. 2013 AAS Meeting 221: 254.11
Ligo, Virgo & Karga Collaboration 2021, astro-ph arXiv:2111.03634
Lupi, A., Haiman, Z. & Volonteri, M. 2021 MNRAS 503, 5046–5060
Madau, P., Rees, M.J., Volonteri, M. et al. 2004 ApJ 604, 484
Madau, P. 2018, MNRAS 480, L43–L47
Madau, P. & Fragos, T. 2017 ApJ 840, 39
Madau P., Meiksin A. & Rees M. J., 1997, ApJ 475, 429
Ma, L. et al. 2021, MNRAS 508, 1973
Mahy, L. H. Sana, H., Abdul-Masih, M. et al. 2020 A&A 634, A118
Malizia, A., Sazonov, S., Bassani, L. et al. 2020, New Astronomy Reviews 90, 101545
Marscher, A. P., Jorstad, S. G., Gomez, J. L. et al. 2002, Nature, 417, 625
McMillan, P. J. 2011, MNRAS, 414, 2446–2457
McNamara, B. R. & Nulsen, P. E. J 2012 NJPh 14, 055023 (40pp)
McQuinn, M. 2012, MNRAS, 426, 1349–1360
Mebane, R.H., Mirocha, J. & Furlanetto, S. R. 2018, MNRAS 479, 4544–4559
Mebane, R. H., Mirocha, J. & Furlanetto, S.R. 2020, MNRAS 493, 1217–1226
Meiksin, A. & Madau, P. 2021, MNRAS 501, 1920–1932
Mellema, G., Koopmans, L., Shukla, H., et al. 2015, Advancing Astrophysics with the Square Kilometre Array (AASKA14), 10
Merloni, A., Heinz, S., & Matteo, T. D. 2003, MNRAS, 345, 1057
Mesinger A., Ferrara A. & Spiegel D. S., 2013, MNRAS, 431, 621
Mesinger, 2019. Understanding the Epoch of Reionization. Challenges and Progress. Astrophysics and Space Science Library vol. 423. Andrei Mesinger editor.
Mezcua, M., Suh, H. & Civano, F. 2019 MNRAS 488, 685–695
Miller-Jones, J. C. A., Tetarenko, A. J., Sivakoff, G. R. et al. 2019, Nature, 569, 374
Miller-Jones, J. C. A., Bahramian, A., Orosz, J.A., et al. 2021, Science 371, 1046–1049
Mirabel, I. F., Dhawan V., Chaty S., et al. 1998 A&A, 330, L9
Mirabel, I. F., Dijkstra, M., Laurent, P., Loeb, A. & Pritchard, J. R. 2011, A&A, 528, A149
Mirabel, I. F., Rodríguez, L. F., Cordier, S. et al. 1992, Nature 358, 215-217
Mirabel, I. F. & Rodríguez, L. F., 1994, Nature 371, 46-48

Mirabel, I. F. & Rodríguez, L. F., 1999, ARAA 37, 409
Mirabel, I. F., & Rodrigues, I. 2003, Science, 300, 1119
Mirocha, J. Mebane, R. H., Furlanetto, S. R. et al. 2018, MNRAS, 478, 5591
Mirocha, J. and Furlanetto, S. R. 2019 MNRAS 483, 1980
Momjian, E., Bañados, E., Carilli, C. L. et al. 2021 ApJ 161, 207
Motta, S. E. Kajava, J. J. E., Giustini, M. et al. 2021, MNRAS 503, 152–161
Muxlow, T., Beswick, R. J., Richards, A. M. S., et al. 2006, Proceedings of the 8th European VLBI Network Symposium,
Natarajan, P., Pacucci, F., Ferrara, A. et al. 2017, ApJ 838, 117
Neeleman, M., Novak, M., Venemans, B. P. et al. 2021 APJ 911, 141
Olejak, A., Belczynski2, K., Bulik, T. et al. 2020 A&A 638, A94
Orosz, J. A., McClintock, J. E., Narayan, R. et al. 2007, Nature 449, 872
Orosz, J. A., Steeghs, D., McClintock, J. E. et al. 2009, ApJ 697, 573–591
Orosz, J. A., McClintock, J. E., Aufdenberg, J. P. et al. 2011, ApJ 742, 840
Paciga G. et al., 2013, MNRAS, 433, 639
Parsons A. R. et al., 2010, AJ, 139, 1468
Philip, L. et al. 2019. Journal of Astronomical Instrumentation, Vol. 08, No. 02, 1950004
Pittard, J. M., Dougherty, S. M., Coker, R. F., et al. 2006 A&A 446, 1001-1019.
Pons, E., McMahon, R. G., Banerji, M., et al. 2020, MNRAS, 491, 3884
Pospelov M., Pradler J., Ruderman J. T. et al. 2018, Phys. Rev. Lett., 121, 031103
Prandoni I. & Seymour N., 2015, Proc. Sci., Revealing the Physics and Evolution of Galaxies and Galaxy Clusters with SKA Continuum Surveys. SISSA, Trieste, PoS#67
Pietrzynski, G., Graczyk, D., Gieren, W., et al. 2013, Nature, 495, 76
Price D. C., Greenhill, L. J., Fialkov, A. et al. 2018, MNRAS, 478, 4193
Pritchard, J.R. & Loeb, A. 2008, Phys. Rev. D 78, 103511
Pritchard, J.R. & Loeb, A. 2010, Phys. Rev. D 82, 023006
Reid, M. J., Menten, K. M., Brunthaler, A., et al. 2019, ApJ, 885, 131
Reines, A. E., Sivakoff, G. R., Johnson, K. E. et al. 2011, Nature 470, 67-68
Reis, I., Fialkov, A. & Barkana, R. 2020 MNRAS 499, 5993–6008
Reis, I., Fialkov, A. & Barkana, R. 2021, MNRAS 506, 5479-5493
Reynolds, T. M., Fraser, M. & Gilmore, M. 2015. MNRAS 453, 2885
Ricotti, M. & Ostriker, J. P. 2004, MNRAS 352, 547–562
Robertson, B. E. 2022, to appear in ARA&A, arXiv:2110.13160
Rodríguez, L. F. & Mirabel, I. F. 2022, in progress
Rodriguez, J.; Grinberg, V.; Laurent, P. et al. 2015 ApJ 807, 17
Rojas-Ruiz, S., Bañados, E., Neeleman, M. et al. 2021, ApJ, 920, 150
Ryden, B. 2016, Introduction to Cosmology, Cambridge, UK: Cambridge University Press
Sana, H., de Mink, S. E., de Koter, A. et al. 2012, Science, 337, 444
Sana, H., Le Bouquin, J.-L., Lacour, S. et al. 2014, ApJS, 215, 15
Sana, H., Ramirez-Tannus, M. C., de Koter, A. et al. 2017, ApJS, 215, 15
Sanders, D. B. & Mirabel, I. F., 1996 ARAA 34, 749-789
Schneider, F. R. N., Sana, H., Evans, C. J. et al. 2021 Science 359 (6371), 69-71.
Scoville, N., Murchikova, L., Walter, F. et al. 2017 ApJ 836, 66
Seiffert, M., Fixsen D. J., Kogut, A. et al. 2011, ApJ 734, 6
Sharma, P. 2108 MNRAS 481, L6
Shimwell T. W. et al., 2019, A&A, 622, A1
Singal, J., Haider, J, Ajello, M. et al. 2018, PASP 130: 036001
Singh S., Subrahmanyan, R., Shankar, N. U. et al. 2018, ApJ, 858, 54
Singh, S., Nambissan T. J., Subrahmanyan, R. et al. 2022/2, NatAs 47S
Smith, B. D. et al. 2018 MNRAS 480, 3762

Soria, R., Pakull, M. W., Motch, C. et al. 2021 MNRAS 501, 1644-1662
Stacy, A., Bromm, V., & Lee, A. T. 2016, MNRAS, 462, 1307
Stirling, A. M., Spencer, R. E., de la Force, C. J. et al. 2001, MNRAS 327, 1273
Subrahmanyan, R. & Cowsik, R. 2013, ApJ 776, 42-50
Sugimura, K., Matsumoto, T. & Hosokawa, T. 2020 ApJ, 892L, 14
Terzian, Y. & Pankonin, V. 1972, ApJ, 174, 293
Tikhonov, N. A., Galazutdinova, O. A. & Karataeva, G. M. 2019, Astrophysical Bulletin 74, 257
Trimble, V. 1968, AJ, 73, 535
Tingay S. J. et al., 2013, PASA, 30, e007
Trushkin S., McCollough M., Nizhelskij, N. et al. 2017 in Galaxies 2017, 5, 86
Turk, M. J., Abel, T., & O'Shea, B. 2009, Science, 325, 601
Tsai, C-W, Eisenhardt, P. R. M., Wu, J. et al. 2015 ApJ 805, 90
Venemans, B. P., Walter, F., Decarli, R. et al. 2017, ApJL 851, L8
Venemans, B. P., Walter, F., Neeleman, M. et al. 2020 ApJ 904, 130
Venumadhav T., Dai L., Kaurov A., Zaldarriaga M., 2018, Phys. Rev. D, 98, 103513
Vinyaikin, E. N. 1993, Astronomy Letters, 19, 369
Vito, F., Brandt, W. N., Yang, G. et al. 2018a MNRAS 473, 2378–2406
Visbal, Haiman, Z. & Bryan, G.L. 2014, MNRAS 445, 1056–1063
Visbal E., Haiman Z., Bryan G. L., 2015, MNRAS, 453, 4456
Visbal E., Haiman Z., Bryan G. L., 2018, MNRAS, 475, 5246
Voytek, T.C., Natarajan, A., Jauregui Garcia, J.M., et al. 2014, ApJL 782, L9.
Volonteri, M. 2012 Science 337, 544-547
Volonteri, M., Silk, J. & Dubus, G. 2015, ApJ, 804, 148
Wang, F., Fan, X., Yang, J. et al. 2020a arXiv:2011.12458v1
Wang, F., Davies, F. B. Yang, J. et al. 2020b ApJ 896, 23
Wang, F., Yang, J., Fang, X. et al. 2021 ApJL 907, L1
Webster, B. L. & Murdin, P. Nature 235, 37–38
Weedman, D. W. 1986, Cambridge Astrophysics Series, Cambridge University Press
Wilms, J., Allen, A. & McCray, R. 2000, ApJ 542, 914
Wise, J. H. et al. 2019 Nature 566, 85-88
Woosley, S. E. 2017, ApJ, 836, 244
Woosley, S. E. & Heger, A. 2021 ApJL 912, L31
Wouthuysen S. A., 1952, AJ, 57, 31
Wright, E. L. 20006 PASP 118, 1711-1715
Wu, X. et al. 2021, MNRAS, 508, 2784–2797
Yang, J., Wang, F., Fan, X. et al. 2020, ApJ, 897, L14
Yan, M., Sadeghpour, H. R. & Dalgarno, A. 1998, ApJ 496, 1044
Yatawatta S. et al., 2013, A&A, 550, A136
Ysard, N. & Lagage, G. 2012, A&A, 547, A53
Zaldarriaga M., Furlanetto S. R., Hernquist L., 2004, ApJ, 608, 622
Zamaninasab, M., Clausen-Brown, E., Savolainen, T. et al. 2014 Nature 510,126
Zapata, L. A., Schmid-Burgk, J., Rodríguez, L. F. et al. 2017, ApJ, 836, 133
Zdziarski, A. A.; Lubinski, P. & Sikora, M. 2012, MNRAS, 423, 663–675
Zdziarski, A. A., Mikolajewska, J. & Belczynski, K. 2013, MNRAS, 429, L104–L108
Zarka, P., Girard, J. N., Tagger, M. et al. 2012 sf2a.conf. 687Z
Zel'dovich, Y.B., Novikov, I. D. 1967 Soviet Astron. 10:602-3
Zubovas, K. & King, A. 2021, MNRAS, 501, 4289-4297